\documentclass{aa}  

\usepackage{graphicx}
\usepackage{xcolor}
\usepackage{indentfirst}

\usepackage{txfonts}
\usepackage{hyperref}
\hypersetup{colorlinks=true,linkcolor=blue,citecolor=blue,filecolor=blue,urlcolor=blue}

\defcitealias{bello2023}{Bello23}
\defcitealias{pass20}{Pass20}
\defcitealias{mar21}{Mar21}
\defcitealias{schw19}{Schw19}
\defcitealias{pass2019}{Pass19}

\begin{document} 

   \title{Using autoencoders and deep transfer learning to determine the stellar parameters of 286 CARMENES M dwarfs \thanks{Table \ref{tab:pars} is only available in electronic form at the CDS via anonymous ftp to \url{cdsarc.u-strasbg.fr (130.79.128.5)} or via \url{http://cdsweb.u-strasbg.fr/cgi-bin/qcat?J/A+A/}}}

   \subtitle{}

   \authorrunning{P.~Mas-Buitrago et al.}
   \titlerunning{Determination of stellar parameters with autoencoders and deep transfer learning}

   \author{P.~Mas-Buitrago\inst{\ref{CSIC-INTA}},
        A.~González-Marcos\inst{\ref{ur}},
        E.~Solano\inst{\ref{CSIC-INTA}},        V.~M.~Passegger\inst{\ref{iac},\ref{ull},\ref{hamb}},
        M.~Cortés-Contreras\inst{\ref{ucm}},
        J.~Ordieres-Mer\'e\inst{\ref{upm}},
        A.~Bello-Garc\'ia\inst{\ref{ovi}},
        J.\,A.~Caballero\inst{\ref{CSIC-INTA}},
        A.~Schweitzer\inst{\ref{hamb}},
        H.~M.~Tabernero\inst{\ref{ucm}},
        D.~Montes\inst{\ref{ucm}},
        and C.~Cifuentes\inst{\ref{CSIC-INTA}}
   }

   \institute{Centro de Astrobiolog\'ia (CAB), CSIC-INTA, Camino Bajo del Castillo s/n, 28692 Villanueva de la Ca\~nada, Madrid, Spain \label{CSIC-INTA} \\
   \email{pmas@cab.inta-csic.es}
        \and
            Departamento de Ingenier\'ia Mec\'anica. Universidad de la Rioja, San Jos\'e de Calasanz 31, 26004 Logro\~no, La Rioja, Spain \label{ur}
        \and
            Instituto de Astrof\'isica de Canarias, c/ V\'ia L\'actea s/n, 38205 La Laguna, Tenerife, Spain \label{iac} 
        \and    
            Departamento de Astrof\'isica, Universidad de La Laguna, 38206 La Laguna, Tenerife, Spain \label{ull}
        \and
            Hamburger Sternwarte, Gojenbergsweg 112, 21029 Hamburg, Germany \label{hamb}
        \and
            Departamento de F{\'i}sica de la Tierra y Astrof{\'i}sica \&
            IPARCOS-UCM (Instituto de F\'{i}sica de Part\'{i}culas y del Cosmos de la UCM), Facultad de Ciencias F{\'i}sicas, Universidad Complutense de Madrid, 28040 Madrid, Spain\label{ucm}
        \and
            Departamento de Ingenier\'ia de Organizaci\'on, Administraci\'on de Empresas y Estad\'istica, Universidad Polit\'ecnica de Madrid, c/~Jos\'e Guti\'errez Abascal 2, 28006 Madrid, Spain \label{upm}
        \and 
            Departamento de Construcci\'on e Ingenier\'ia de Fabricaci\'on, Universidad de Oviedo, Pedro Puig Adam, Sede Departamental Oeste, M\'odulo 7, 1$^a$ planta, 33203 Gij\'on, Spain \label{ovi}
        }

   \date{Received 05 March 2024 / Accepted 02 May 2024}

 
  \abstract
   {Deep learning (DL) techniques are a promising approach among the set of methods used in the ever-challenging determination of stellar parameters in M dwarfs. In this context, transfer learning could play an important role in mitigating uncertainties in the results due to the synthetic gap (i.e. difference in feature distributions between observed and synthetic data).}
   {We propose a feature-based deep transfer learning (DTL) approach based on autoencoders to determine stellar parameters from high-resolution spectra. Using this methodology, we provide new estimations for the effective temperature, surface gravity, metallicity, and projected rotational velocity for 286 M dwarfs observed by the CARMENES survey.}
   {Using autoencoder architectures, we projected synthetic PHOENIX-ACES spectra and observed CARMENES spectra onto a new feature space of lower dimensionality in which the differences between the two domains are reduced. We used this low-dimensional new feature space as input for a convolutional neural network to obtain the stellar parameter determinations.}
   {We performed an extensive analysis of our estimated stellar parameters, ranging from 3050 to 4300\,K, 4.7 to 5.1\,dex, and $-$0.53 to 0.25\,dex for $T_{\rm eff}$, log \textit{g}, and [Fe/H], respectively. Our results are broadly consistent with those of recent studies using CARMENES data, with a systematic deviation in our $T_{\rm eff}$ scale towards hotter values for estimations above 3\,750\,K. Furthermore, our methodology mitigates the deviations in metallicity found in previous DL techniques due to the synthetic gap.}
   {We consolidated a DTL-based methodology to determine stellar parameters in M dwarfs from synthetic spectra, with no need for high-quality measurements involved in the knowledge transfer. These results suggest the great potential of DTL to mitigate the differences in feature distributions between the observations and the PHOENIX-ACES spectra.}


   \maketitle


\section{Introduction}

Low-mass dwarfs are the most common type of stars in the Galaxy, constituting approximately 70\% of the stellar population  \citep{henry1994,reid1995,reyle2021}. In particular, M dwarfs, which are smaller, cooler, and fainter than Sun-like stars are of great importance in the study of exoplanets because of their prevalence, longevity, and proximity. Their small size and lower luminosity make it easier to detect Earth-sized planets in their habitable zones. 
As a result, several programs have been established with the goal of identifying potentially habitable planets orbiting M dwarfs. Notable examples include ground-based instruments like the Echelle Spectrograph for Rocky Exoplanet and Stable Spectroscopic Observations \citep[ESPRESSO,][]{pepe21} and its predecessor, the High-Accuracy Radial velocity Planet Searcher \citep[HARPS,][]{mayor2003,bonfils13}, or the Calar Alto high-Resolution search for M dwarfs with Exoearths with Near-infrared and optical Echelle Spectrographs \citep[CARMENES,][]{Quirrenbach16,Quirrenbach20}.

The precise determination of the stellar parameters of M dwarfs is crucial to improve our understanding of planetary formation and evolution, which depends fundamentally on the thorough characterisation of their host stars \citep{cifuentes2020}. However, well-established photometric and spectroscopic methods for determining these parameters encounter particular challenges, mainly due to the inherent faintness of M dwarfs and their frequent manifestation of strong stellar activity. Specifically for spectroscopic analyses, establishing the spectral continuum can be a difficult task.
Despite these problems, numerous efforts have been devoted to estimating photospheric parameters in M dwarfs, including effective temperature ($T_{\rm eff}$), surface gravity (log \textit{g}), and metallicity ([M/H]). Several methods have proven successful in inferring these parameters, such as fitting synthetic spectra, as in  \citet[][hereafter \citetalias{pass2019}]{pass2019} and  \citet[][hereafter \citetalias{mar21}]{mar21}, pseudo-equivalent widths (pEWs) \citep[e.g.][]{Mann2013,Mann2014,Neves2014}, 
spectral indices \citep[e.g.][]{RojasAyala2010,RojasAyala2012}, empirical calibrations \citep[e.g.][]{casagrande08,Neves2012}, interferometry \citep[e.g.][]{Boyajian2012,Rabus2019}, and machine learning  \citep[e.g.][hereafter \citetalias{pass20}]{Antoniadis2020,pass20}.

The approaches based on pEWs, measurements of the strength of absorption lines in a spectrum, and spectral indices, calculated from carefully chosen spectral regions --and  often derived from absorption lines or bands--, leverage their sensitivity and correlation with stellar parameters (mainly, $T_{\rm eff}$ and [Fe/H]). As a recent example of these approaches, \cite{Khata2020} determined $T_{\rm eff}$ and  metallicities, among other parameters, for 53 M dwarfs using $H$- and $K$-band pEWs and H$_{2}$O indices.
Another approach relies on empirical calibrations based on observations of M dwarfs that have an F, G, or K binary companion with known metallicity. This is grounded in the idea that the metallicity of an M dwarf is comparable to that of the hotter primary star, assuming the system originated from the same proto-stellar cloud \citep{Neves2012,montes2018,duque24}. For example, \cite{Rodriguez2019} employed the relationships of \cite{Newton2015} and \cite{Mann2013b} to derive $T_{\rm eff}$ and metallicity, respectively, from moderate-resolution spectra of 35 M dwarfs from the \textit{K2} mission. 
Numerous spectral indices have also been empirically calibrated. For instance, \cite{Veyette2017} determined $T_{\rm eff}$, [Fe/H], and [Ti/H] from high-resolution \textit{Y}-band spectra of 29 M dwarfs by combining spectral synthesis with empirically calibrated indices and pEWs using FGK+M systems \citep{bonfils2005,Mann2013}.

Interferometric measurements have also proven useful for deriving index-based calibrations for $T_{\rm eff}$ \citep{Mann2013b}, performing empirical calibrations for $T_{\rm eff}$ \citep{maldonado2015,Newton2015}, or determining $T_{\rm eff}$ from interferometric observations in combination with parallaxes and bolometric fluxes \citep{Boyajian2012,vonBraun2014,Rabus2019}. However, their application is limited to a relatively small number of stars due to the requirement that they must be bright and nearby.

The fitting of synthetic spectra relies on a minimisation algorithm to find the synthetic spectrum that best matches the observed spectrum. Variations exist in terms of the synthetic grid employed (e.g. BT-Settl, PHOENIX-ACES, MARCS), using high or low spectral resolution, and the number and wavelength of features selected for comparison.
For example, the BT-Settl models \citep{Allard2012,Allard2013} were used by \cite{GaidosMann2014} and \cite{mann2015} to derive $T_{\rm eff}$ values for M dwarfs with low-resolution visible SNIFS (Supernova Integral Field Spectrograph) spectra, and by \cite{Rajpurohit2018} to compute $T_{\rm eff}$, log \textit{g}, and [Fe/H] for 292 M dwarfs using high-resolution CARMENES spectra \citep{reiners2018}. \cite{Kuznetsov2019} applied BT-Settl models to intermediate-resolution spectra from the visible arm of VLT/X-shooter \citep[intermediate resolution, high-efficiency spectrograph,][]{Vernet2011} to determine $T_{\rm eff}$, log \textit{g}, [Fe/H], and $v\sin{i}$ for 153 M dwarfs. More recently, \cite{Hejazi2020} derived $T_{\rm eff}$, $\log{g}$, 
metallicity [M/H], and alpha-enhancement [$\alpha$/Fe] of 1\,544 M dwarfs and subdwarfs from low- to medium-resolution spectra collected at the Michigan-Dartmouth-MIT observatory, Lick Observatory, Kitt Peak National Observatory, and Cerro Tololo Interamerican Observatory. Additionally, \citetalias{mar21} determined $T_{\rm eff}$, log \textit{g}, and [Fe/H] for a sample of 343 M dwarfs observed with CARMENES using a Bayesian implementation of the spectral synthesis technique, the \texttt{SteParSyn}\footnote{\url{https://github.com/hmtabernero/SteParSyn}} code \citep{Tabernero2022}.

Based on the PHOENIX-ACES library \citep{Husser2013}, \cite{Birky2017} derived $T_{\rm eff}$, log \textit{g}, and [Fe/H] for late-M and early-L dwarfs from high-resolution near-infrared APOGEE spectra \citep{Wilson2010}. Similarly, \cite{pass18} and  \citet[][hereafter \citetalias{schw19}]{schw19} determined these parameters for M dwarfs observed with CARMENES in the visible wavelength region. Building upon these works,  \citetalias{pass2019} extended the analysis by determining $T_{\rm eff}$, log \textit{g}, and [Fe/H]  not only from the visible range covered with CARMENES but also from the near-infrared and the combination of visible and near-infrared data. The comparison conducted in \citetalias{pass2019} led to the conclusion that utilising both spectral ranges for parameter determination maximises the amount of available spectral information while minimising possible effects caused by imperfect modelling.
The MARCS model atmospheres \citep{Gustafsson2008} have also been employed to compute photospheric parameters. For instance, in a recent study by \cite{Souto2020}, $T_{\rm eff}$, log \textit{g}, and [Fe/H] were determined for 21 M dwarf mid-resolution APOGEE \textit{H}-band spectra using MARCS models and the {\ttfamily turbospectrum} code \citep{Plez2012} through the {\ttfamily bacchus} wrapper \citep{Masseron2016}. 
Similarly, \cite{Sarmento2021} derived $T_{\rm eff}$, log \textit{g}, [M/H], and microturbulent velocity $v_{\rm mic}$ for 313 M dwarfs from APOGEE \textit{H}-band spectra using MARCS models, {\ttfamily turbospectrum}, and {\ttfamily iSpec} python code \citep{BlancoCuaresma2014}.

As large surveys release extensive databases containing thousands of stars, there is a need for flexible and automated methods capable of handling vast amounts of data to infer stellar atmospheric parameters. In this sense, machine learning (ML) techniques have also been used for determining photospheric parameters for M dwarfs from stellar spectra. For example, \cite{Sarro2018} proposed an automated procedure based on genetic algorithms to identify pEWs and integrated flux ratios from BT-Settl models that yield good estimations of $T_{\rm eff}$, log \textit{g}, and [M/H] for spectra from the NASA Infrared Telescope Facility (IRTF). Also based on pEWs, \cite{Antoniadis2020} present an ML tool, named {\ttfamily ODUSSEAS}, to derive $T_{\rm eff}$ and [Fe/H] of M dwarf stars from 1D spectra for different resolutions. In \cite{Birky2020}, {\ttfamily The Cannon} \citep{Ness2015,Casey2016}, a data-driven spectral-modelling and parameter-inference framework, is used to estimate $T_{\rm eff}$ and [Fe/H] for 5\,875 M dwarfs in the APOGEE \citep{Abolfathi2018} and {\it Gaia} DR2 \citep{GaiaDR2} surveys. Using the Stellar LAbel Machine \citep[SLAM,][]{Zhang2020}, \cite{Li2021} trained a model with APOGEE stellar labels and synthetic spectra from the BT-Settl model, resulting in the determination of $T_{\rm eff}$ and [M/H] for M dwarfs from the LAMOST DR6\footnote{\url{http://dr6.lamost.org/}} catalogue.

This study extends previous works on applying deep learning (DL) to predict stellar parameters from high-resolution spectra observed with CARMENES. \citetalias{pass20} presented a DL approach where convolutional neural networks (CNNs) were trained on synthetic PHOENIX-ACES models to estimate $T_{\rm eff}$, log \textit{g}, [M/H], and $v\sin{i}$ for 50 M dwarfs observed with CARMENES. After a thorough analysis of their methodology, in which different architectures and spectral windows were tested, they found that all DL models were able to estimate stellar parameters from synthetic spectra in a precise and accurate way. However, when testing these models on the CARMENES spectra, they found significant deviations for the metallicity because of the synthetic gap \citep{fabbro2018,Tabernero2022}, which is the difference in feature distributions between synthetic and observed data.
In a more recent study, \citet[][hereafter \citetalias{bello2023}]{bello2023} employed a deep transfer learning (DTL) approach to mitigate the uncertainties associated with the synthetic gap (see their Figs. 1 and 2). Following the training of DL models on a large set of synthetic spectra from the PHOENIX-ACES model, the models underwent fine-tuning based on external knowledge about stellar parameters. This external knowledge included 14 stars from the CARMENES survey with interferometric angular diameters measured by \citet{Boyajian2012}, \citet{vonBraun2014}, and references therein. Additionally, it was supplemented with five mid-to-late M dwarf stars from \citet{passegger2022}. They achieved the determination of new $T_{\rm eff}$ and [M/H] values for 286 M dwarfs from the CARMENES survey, and although this approach improved the estimation of $T_{\rm eff}$ and [M/H] for M dwarfs from high-resolution spectra obtained with CARMENES, the lack of sufficiently large number of reference stars to transfer knowledge is a limitation for the technique. If the reference dataset is limited in size, diversity, or representation across the parameter space, the models may not generalise well to a broader range of M dwarfs.

In this work, we present a novel transfer learning approach for estimating photospheric parameters in M dwarfs based on their stellar spectra. The primary goal of the proposed method is to address the aforementioned limitation identified by \citetalias{bello2023} by eliminating the requirement for interferometric values in the knowledge transfer process.
To achieve this, instead of employing a model-based transfer learning approach, as in \citetalias{bello2023}, where the transferred knowledge is encoded into model parameters, priors or model architectures, we propose a feature-based transfer learning. In this approach, the knowledge to be transferred can be considered as the learned feature representation. The idea is to learn a `good' feature representation so that, by projecting data onto the new representation, the differences between domains (source and target, i.e. synthetic and observed spectra in our case) can be reduced. This allows the source domain labelled data (synthetic spectra with known parameters) to be used to train a precise model for the target domain constituted by the observed spectra \citep{yang2020}.

In Section \ref{sec:data}, we provide details on the CARMENES sample and the PHOENIX-ACES synthetic model grid used in this study. The proposed methodology, based on autoencoders and transfer learning, is outlined in Section \ref{sec:methodology}. The derived stellar atmospheric parameters are then analysed and compared with existing literature in Section \ref{sec:results}. Finally, Section \ref{sec:conclusions} summarises the main conclusions of this work.


\section{Data} \label{sec:data}

The proposed approach was tested using the same sample spectra as \citetalias{pass2019}. This sample, listed in their Table B.1, comprise 282 M dwarfs observed with CARMENES. Additionally, four more stars from an independent interferometric sample, as described by \citetalias{bello2023}, were included.

CARMENES is installed at the Calar Alto Observatory, located in Spain, and stands as one of the leading instruments in the quest for searching for Earth-like planets within the habitable zones around M dwarfs. It comprises two separate spectrographs: one for the visible (VIS) wavelength range (from 520 to 960\,nm) and the other for the near-infrared (NIR) range (from 960 to 1710\,nm), each offering high-spectral resolutions of R\,$\approx$\,94\,600 and 80\,500, respectively \citep{Quirrenbach20,reiners2018}. 

A detailed description of the data reduction procedure is available in \citet{Zechmeister14}, \citet{Caballero2016}, and \citetalias{pass2019}.
Similar to the latter, we used a high signal-to-noise (S/N) template spectra for each star. These templates are generated as byproducts of the CARMENES radial-velocity pipeline, known as {\tt serval} 
\citep[SpEctrum Radial Velocity AnaLyser;][]{Zechmeister2018}. In the standard data flow, the code constructs a template for each target star from a minimum of five individual spectra to derive the radial velocities through least-square fitting to the template. 
The S/N of the observed CARMENES sample used in this work was above 150. Concerning the wavelength window, we adopted the range 8\,800--8\,835\,\AA, consistent with \citetalias{bello2023}, as this window displayed the smallest mean squared error among all the investigated windows in \citetalias{pass20}.

To train the neural network models, we utilised the PHOENIX-ACES spectra library\footnote{\url{https://phoenix.astro.physik.uni-goettingen.de/}} \citep{Husser2013}. This library is chosen for its consideration of spectral features present in cool dwarfs. Furthermore, the use of synthetic models enables the generation of a large number of spectra with known parameters, eliminating the need for limited samples of observations with well-known stellar parameters. We used the same PHOENIX-ACES grid as in previous works (\citetalias{pass20}; \citetalias{bello2023}), which was generated by linearly interpolating between the existing grid points using {\ttfamily pyterpol} \citep{Nemravov2016}. The complete dataset contains a grid of 449\,806 synthetic high-resolution spectra between 8\,800\,{\AA} and 8\,835\,{\AA} with $T_{\rm eff}$ between 2\,300 and 4\,500\,K (step 25\,K), log \textit{g} between 4.2 and 5.5\,dex (step 0.1\,dex), [M/H] between -1.0 and 0.8\,dex (step 0.1\,dex), and $v\sin{i}$ between 1.5 and 60.0\,km\,s$^{-1}$ (with a variable step of 0.5, 1.0, 2.0 or 5.0; see Table 1 in \citetalias{pass20}). A degeneracy between $T_{\rm eff}$, log \textit{g}, and [Fe/H] was described by \citet{pass18}, who found exceptionally high values of log \textit{g} and [Fe/H] for well-fitting PHOENIX-ACES models. This degeneracy was further underscored by \citetalias{pass2019} and \citetalias{pass20} during the application of DL models to the observed CARMENES spectra, and the latter imposed additional constraints to the grid leveraging the PARSEC v1.2S evolutionary models \citep{bressan2012,Chen2014,Chen2015,Tang2014}. Degeneracies between stellar parameters are often found when fitting synthetic spectra, and some authors have explored several ways to help break them \citep{buzzoni2001,brewer2015}. The refinement performed by \citetalias{pass20} aimed to exclude parameter combinations for M dwarfs that do not fit the main sequence, as discussed in Section 4.2 of their work. Notably, \citetalias{pass20} demonstrated that the imposition of these constraints on the synthetic model grid used in the training of the DL models is capable of breaking the observed parameter degeneracy. After applying these restrictions, the grid includes 22\,933 PHOENIX-ACES spectra.

Due to the negligible presence of telluric features in the investigated range, telluric correction was not applied to the VIS spectra.
For normalisation, we employed the Gaussian Inflection Spline Interpolation Continuum ({\tt GISIC}\footnote{\url{https://pypi.org/project/GISIC/}}), the same method and routine used by \citetalias{pass20} and developed by D.\,D.~Whitten, designed for spectra with strong molecular features. 
Following the same approach as \citetalias{bello2023}, we applied this procedure to both observed and synthetic spectra within the spectral window 8800--8835\,\AA{} with an additional 5\,\AA{} on each side to mitigate potential edge effects. Moreover, the observed spectra underwent radial velocity correction to align with the rest frame of the synthetic spectra, achieved through cross-correlation \citep[\texttt{crosscorrRV} from PyAstronomy,][]{Czesla2019} between a PHOENIX model spectrum and the observed spectrum. To ensure a universal wavelength grid, essential for applying the proposed method, the wavelength grid of the observed spectra was linearly interpolated with the grid of the synthetic spectra.

\begin{figure}
	\includegraphics[width=\columnwidth]{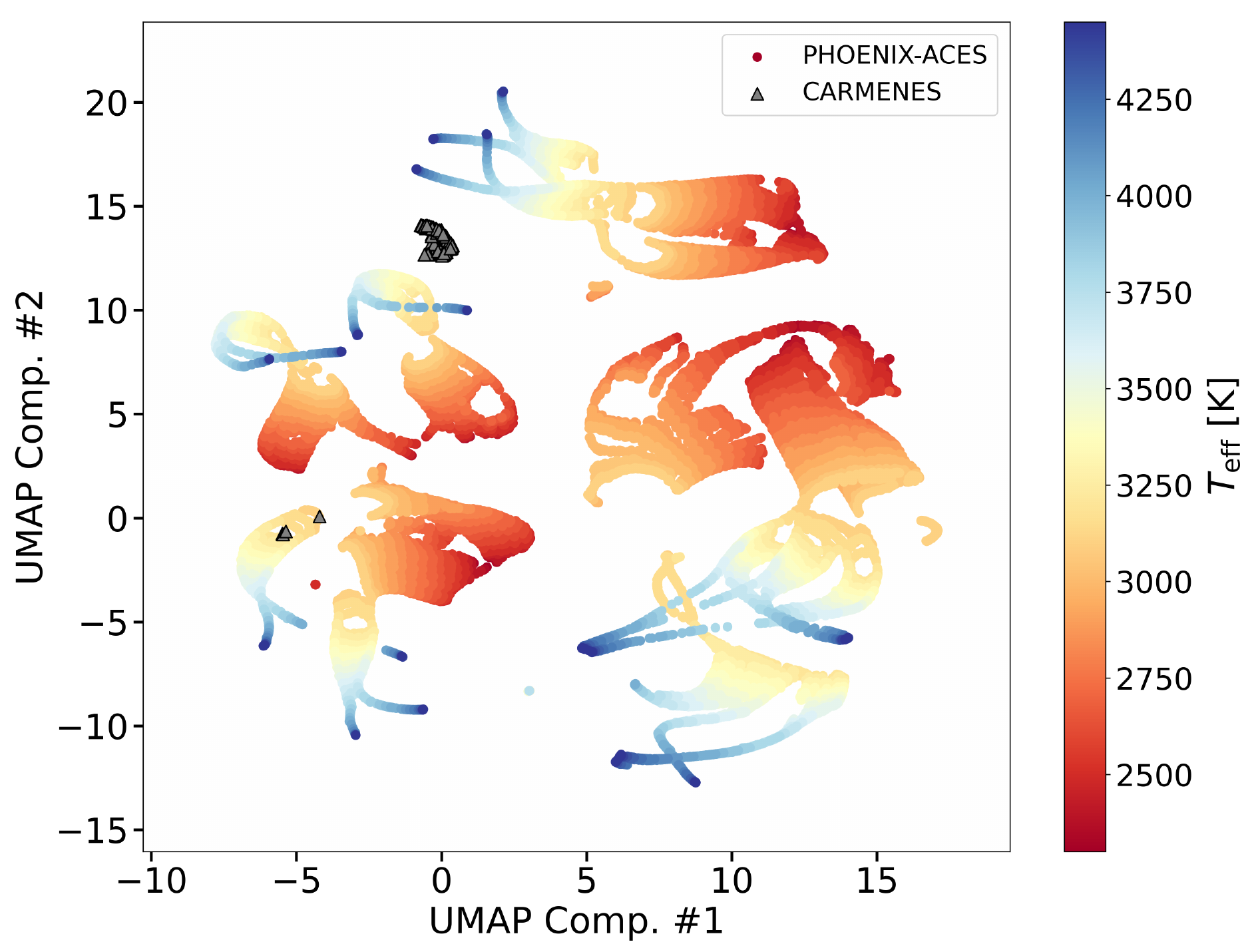}
    \caption{Two-dimensional UMAP projection of PHOENIX-ACES (dots colour-coded by $T_{\rm eff}$) and CARMENES (grey triangles) spectra from the 8\,800--8\,835\,$\AA$ window. Almost all CARMENES spectra are isolated from the PHOENIX-ACES family feature space.}
    \label{fig:umap_flux}
\end{figure}

In spite of the performed spectra preparation, differences in the feature distributions of the synthetic and observed sets of spectra (i.e. synthetic gap) were identified. We used the Uniform Manifold Approximation and Projection \citep[UMAP;][]{mcinnes2018umap}, with a metric that considers the correlation between the spectra, to project the high-dimensional input space (3\,500 flux values for each spectrum) into a two-dimensional space while preserving inter-distances. As shown in Fig. \ref{fig:umap_flux}, akin to \citetalias{pass20} and \citetalias{bello2023}, most of the CARMENES spectra (grey triangles) do not align precisely within the synthetic spectra (colour-coded dots). Thus, a transfer learning approach appears appropriate to extend the applicability of the regression models trained with the synthetic spectra to the observed spectra. 

\section{Methodology} \label{sec:methodology}

The DTL approach proposed in this paper can be summarised as follows. Initially, we extract a low-dimensional representation of synthetic spectra based on the PHOENIX-ACES library using autoencoders (AEs), a special kind of neural network initially proposed for dimensionality reduction \citep{hinton2006}. Then, the knowledge transfer process is performed by fine-tuning these AEs with high-resolution spectra observed with the CARMENES instrument. It must be noted that no stellar parameters were used during this re-training. With the low-dimensional representations of the synthetic spectra resulting from the initial step, we trained CNNs. Finally, using these CNNs, we estimated the stellar parameters ($T_{\rm eff}$, log \textit{g}, [M/H], and $v\sin{i}$) for 286 CARMENES M dwarfs by using their low-dimensional representations obtained after the fine-tuning step.

\subsection{Feature extraction using an autoencoder} \label{sec:ac}

In this study, we explore unsupervised feature extraction from stellar spectra using AEs to facilitate feature-based transfer learning and leverage the new representations for estimating photospheric parameters. 
Belonging to representation learning --a subfield of machine learning--, AEs have the capability to capture the underlying factors hidden in the observed data \citep{Bengio13,Goodfellow16}. 
They have been succesfully used in various astrophysical applications, including unsupervised feature learning from galaxy spectral energy distribution \citep{FronteraPons17}, learning of non-linear representations from rest-frame spectroscopic data for redshift estimation \citep{FronteraPons19}, galaxy classification \citep{Cheng21}, astrophysical component separation \citep{Milosevic21}, reconstruction of missing magnitudes from observed objects before classifying them into stars, galaxies, and quasars \citep{Khramtsov21}, and telluric correction \citep{Kjarsgaard23}. 
In addition, some authors have used AEs to estimate stellar atmospheric parameters from spectra \citep{Yang2015,Li2017}. However, their approach is different from our proposal since the training of the models was performed in a supervised manner: spectra from SDSS/SEGUE DR7 \citep{Abazajian2009} were used, and $T_{\rm eff}$, log \textit{g}, and [Fe/H] were obtained from the SDSS/SEGUE Spectroscopic Parameter Pipeline \citep[SSPP;][]{SSPP01,SSPP02,SSPP03,SSPP04} for stars in the temperature range 4088-9747 K (earlier than our CARMENES targets). 
In our case, we are interested in the use of AEs to enable transfer learning, as representation learning enables the transfer of knowledge when there are features useful for different settings or tasks that correspond to underlying factors appearing in more than one setting \citep{Goodfellow16}.

\begin{figure*}
\centering
	\includegraphics[width=\linewidth]{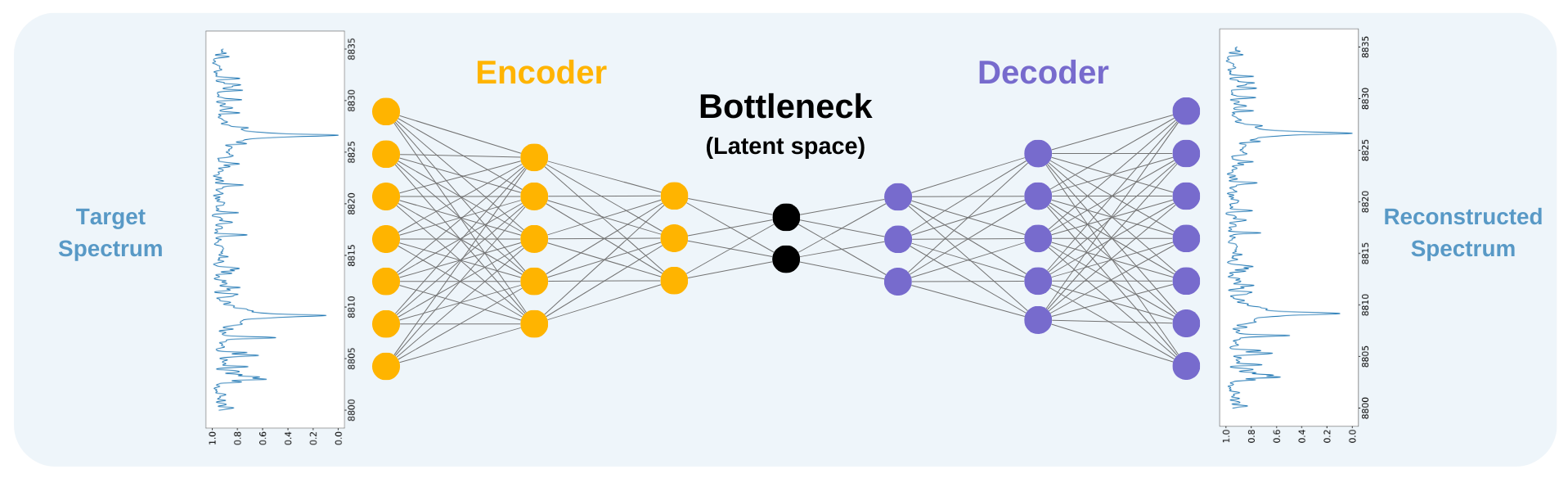}
    \caption{Schematic representation of the AE architecture used in this work.}
    \label{fig:ac_info}
\end{figure*}

The rationale behind the first step of our methodology is to find a meaningful low-dimensional representation, referred to as the latent space, of the synthetic spectra. To accomplish this, we employed an AE, which consists of an `encoder' trained to transform the high-dimensional spectrum into a low-dimensional code, and a `decoder' trained to reconstruct the original spectrum as accurate as possible from its lower-dimensional latent space (see Fig. \ref{fig:ac_info}).

First, we divided the grid of synthetic spectra into a training set (70\,\%) and a test set (30\,\%). We considered multiple AE architectures, developing a python code to create a flexible AE structure. The number of neurons on each layer, the L1 regularisation term for the dense layers (used to prevent overfitting), and the learning rate for the Adam optimisation \citep[a computationally efficient stochastic gradient descent method,][]{adam} were passed as parameters. For this code, we relied on the \texttt{Keras}\footnote{\url{https://keras.io/about/}} \citep{keras} deep learning API, which runs on top of the \texttt{Tensorflow}\footnote{\url{https://www.tensorflow.org/}} \citep{tensorflow} machine learning platform. Next, we created a grid for these hyperparameters and performed an exhaustive search using the \texttt{GridSearchCV} class from the \texttt{scikit-learn}\footnote{\url{https://scikit-learn.org/stable/}} package, which optimises the hyperparameters of an estimator through k-fold cross-validation, using any scoring metric to evaluate the model. In our case, we used 4-fold cross-validation and the mean squared error between the reconstructed and the original validation data as the scoring metric. To integrate our python code into a \texttt{scikit-learn} workflow, we used the \texttt{KerasRegressor} wrapper from the \texttt{scikeras}\footnote{\url{https://adriangb.com/scikeras/stable/}} python package.

After this search for the best hyperparameter combinations, we only kept those with a mean cross-validation score below the median, evaluated using the entire grid. We trained an AE for each of these architectures, adding a contractive regularisation term in the loss function, consisting of the squared Frobenius norm of the Jacobian matrix of the encoder activations with respect to the input:

\begin{equation}
    \left \|\,J_{f}\,(x)\, \right \|_{F}^{2}=\sum_{ij}\left ( \frac{\partial h_{j}\,(x)}{\partial x_i} \right)^2,
	\label{eq:contractive_loss}
\end{equation}

\noindent where $f$ represents the encoding function that maps the input $x$ to the hidden representation $h$. The main idea of contractive AEs is to make the feature extraction more robust to small perturbations in the training data. In the overall loss function optimisation, the trade-off between the reconstruction and the L1 regularisation terms will retain the important variations in the latent space for the reconstruction of the input \citep{rifai2011}.

\begin{figure*}
\centering
	\includegraphics[width=\linewidth]{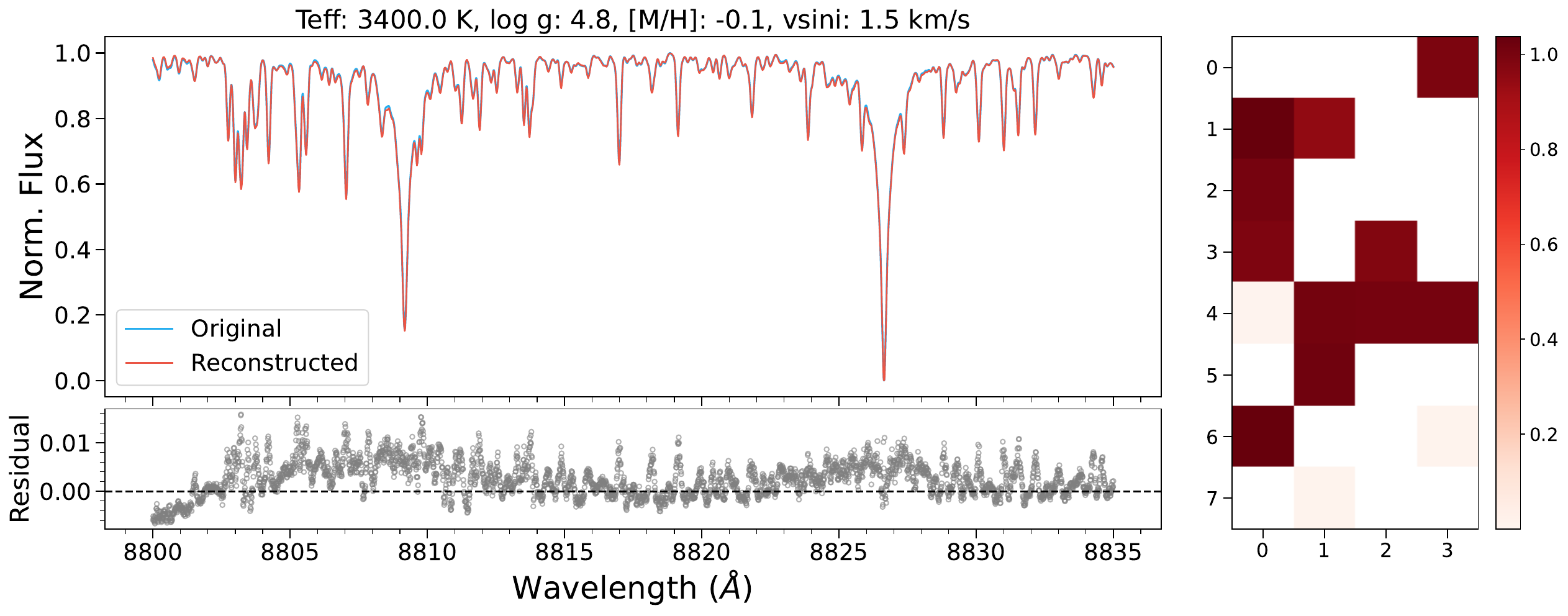}
    \caption{Reconstructed spectrum (\textit{left}) and latent representation (\textit{right}) of a PHOENIX-ACES synthetic spectrum for one of the trained AEs. \textit{Left panel:} comparison of the original (blue) and reconstructed (red) spectrum. Both spectra overlap as they are almost similar. The title shows the stellar parameters of the synthetic spectrum. Reconstruction residuals (original$-$reconstructed) are shown in the \textit{bottom panel}. \textit{Right panel:} 32-dimensional latent space of the input spectrum obtained by the encoder, reshaped to a 8$\times$4 matrix only for a better visibility. The colour scale indicates the strength of the features. The decoder uses this compressed representation to obtain the reconstructed spectrum.}
    \label{fig:ph_rec}
\end{figure*}

We only kept the AEs with a learning rate equal to 0.0001, as we found that some of them with a higher learning rate were not able to converge properly, leading to a poor latent representation of the spectra. With this, we ended up with 26 final AE architectures and evaluated them on the test set, obtaining mean squared reconstruction errors $\sim 5\cdot10^{-5}$. Fig. \ref{fig:ph_rec} shows the reconstruction and the latent space of a PHOENIX-ACES synthetic spectrum for one of the AEs. Using the encoder networks of the AEs, we obtained 26 sets (one for each AE) of 32-dimensional compressed representations for the grid of synthetic spectra.

\subsection{Deep transfer learning} \label{sec:dtl}

The dependence of DL algorithms on massive training data is a crucial hurdle to overcome when a research scenario requires labelled data. In some fields, such as astrophysics, building a large, annotated data set can be incredibly complex and expensive. A straightforward and widely used solution to this problem is the use of synthetic data to train the DL models, but this may include a systematic error in the methodology if the synthetic gap (see Section \ref{sec:data}) is significant, as is the case in this work.

Transfer learning (TL) plays a key role in solving the above problems, as it allows knowledge to be transferred from a rich source domain to a related but not identical target domain. The transition from TL to deep transfer learning (DTL), with incomplete DTL as an intermediate stage \citep[deep neural networks are only used as feature extractors in TL models;][]{yu2022}, came with the integration of DL techniques into the TL paradigm.

In the context of TL, a domain can be represented as $D=\left \{\mathcal{X},P(X)  \right \}$, where $\mathcal{X}$ denotes a feature space and $P(X)$ represents the marginal probability distribution for $X=\left \{x_1,...,x_n \right \} \in \mathcal{X}$. Also,  a task can be represented as $T=\left \{Y, f(\cdot) \right \}$, where $Y$ denotes a label space and $f(\cdot)$ is a predictive function. According to the definition provided by \citet{pan2010}, given a source domain $D_{\rm S}$ and task $T_{\mathrm{S}}$, and a target domain $D_{\mathrm{T}}$ and task $T_{\mathrm{T}}$, TL aims to enhance the performance of a predictive function $f_{\mathrm{T}}(\cdot)$ in $D_{\mathrm{T}}$, using the knowledge available in $D_{\mathrm{S}}$ and $T_{\mathrm{S}}$, where $D_{\mathrm{S}}\neq D_{\mathrm{T}}$ and/or $T_{\mathrm{S}}\neq T_{\mathrm{T}}$. In our work, the source domain is represented by the grid of synthetic PHOENIX-ACES spectra, while the target domain is built from the 286 CARMENES observed spectra. Moreover, the predictive function is defined as the encoder network of the AE architecture, responsible for compressing the input spectra into the low-dimensional latent representation.

The purpose of this step in the methodology is to adopt a DTL-based strategy, in particular the fine-tuning approach \citep{brian2016,yosinski2014}, using the AE architectures we already trained in the source domain to obtain a meaningful low-dimensional latent representation of our data-poor target domain. In this process, we kept the weights frozen in all encoder layers until the last one, leaving only the deepest encoder layer, the bottleneck (i.e. the latent space or compressed representation of the spectrum, as illustrated in Fig. \ref{fig:ac_info}), and the decoder network to be re-trained. The motivation for keeping the lower layers frozen is to prevent generic learning from being overwritten, thus preserving the knowledge acquired by the network to recognise relevant spectral features, while the more specific features are tailored to the target domain \citep{sadr2020}.

\citet{pan2018} already explored the possibility of finding a low-dimensional latent space in which source and domain data are close to each other, and using it as a bridge to transfer the knowledge from the labelled source domain to the unlabelled target domain. In our case, the ultimate goal of this process is to find a low-dimensional representation of the observed spectra that is closer to the synthetic latent representation than in the initial high-dimensional space of the spectra (see Fig. \ref{fig:umap_flux}). Furthermore, we want for these target representations to be as meaningful as possible, since we intend to use them later as a starting point for estimating the stellar parameters.

First, we divided the target set of 286 CARMENES spectra into a training set (80\,\%) and a test set (20\,\%), with the latter being used to assess the reconstruction error across the target domain. Then, we fine-tuned the 26 AE architectures, following the process explained above, obtaining mean squared reconstruction errors $\sim4\cdot 10^{-4}$ on the test set, in contrast to the reconstruction errors ($\sim3\cdot 10^{-3}$) obtained on the CARMENES set using the AEs pre-trained on the PHOENIX-ACES spectra. It must be noted that no stellar parameters were used during this re-training.

\begin{figure}
	\includegraphics[width=\columnwidth]{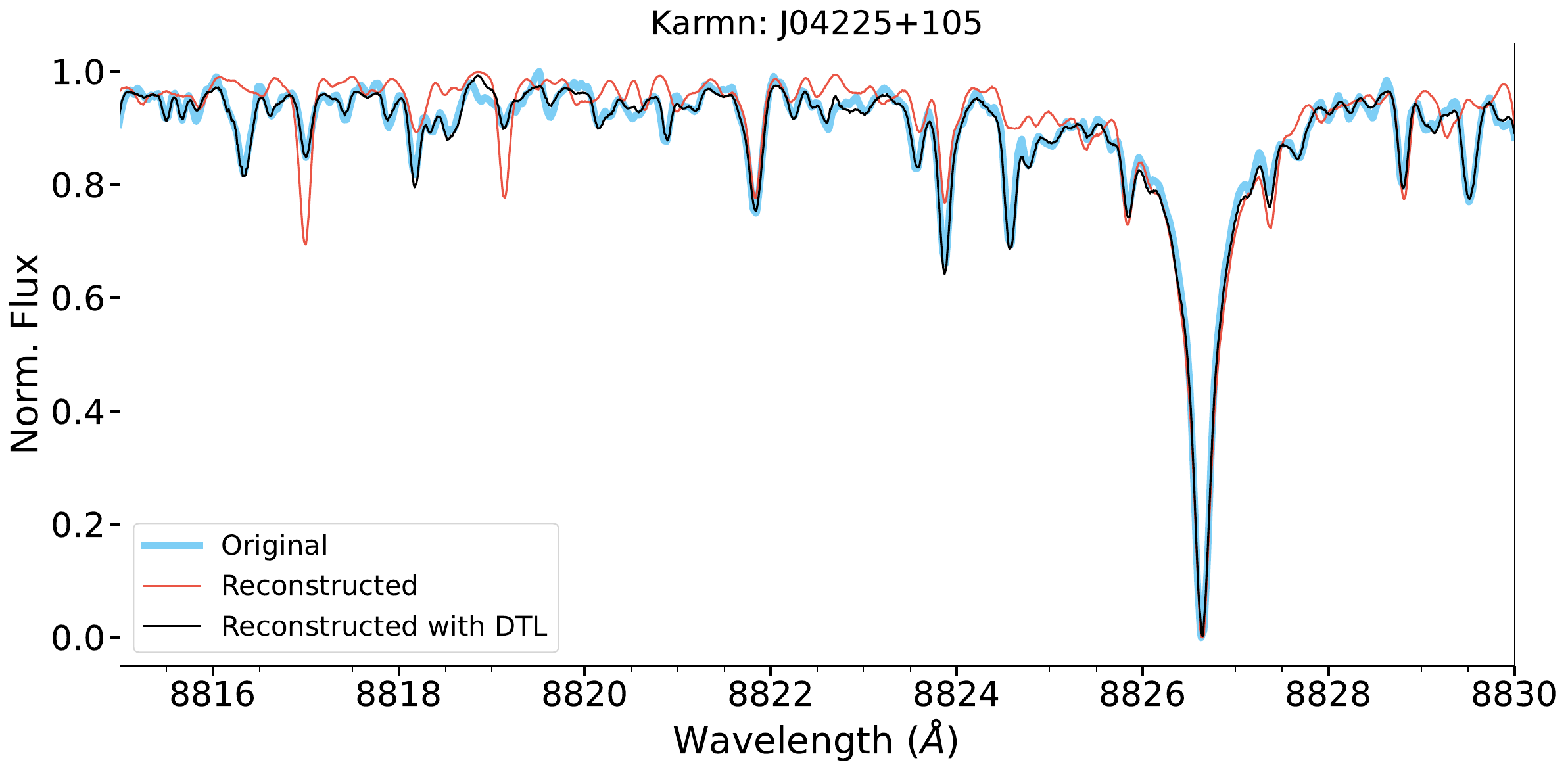}
    \caption{Original (blue) vs. reconstructed CARMENES spectrum for LSPM J0422+1031 (Karmn J04225+105, M3.5\,V). The Figure only shows a section of the spectrum for better visibility, with the unique purpose of emphasising how the reconstruction after fine-tuning (black) captures much more detailed spectral features than the reconstruction with the initial training (red).}
    \label{fig:vsdtl}
\end{figure}

Fig. \ref{fig:vsdtl} illustrates the importance of this step for the AE to effectively adapt to our specific target domain, ensuring that the compressed representations provided by the fine-tuned encoders will be more meaningful than those we would have obtained with the initial training. Using these fine-tuned encoder networks, we obtained the final 26 sets of 32-dimensional representations for the observed CARMENES spectra.

\begin{figure}
	\includegraphics[width=\columnwidth]{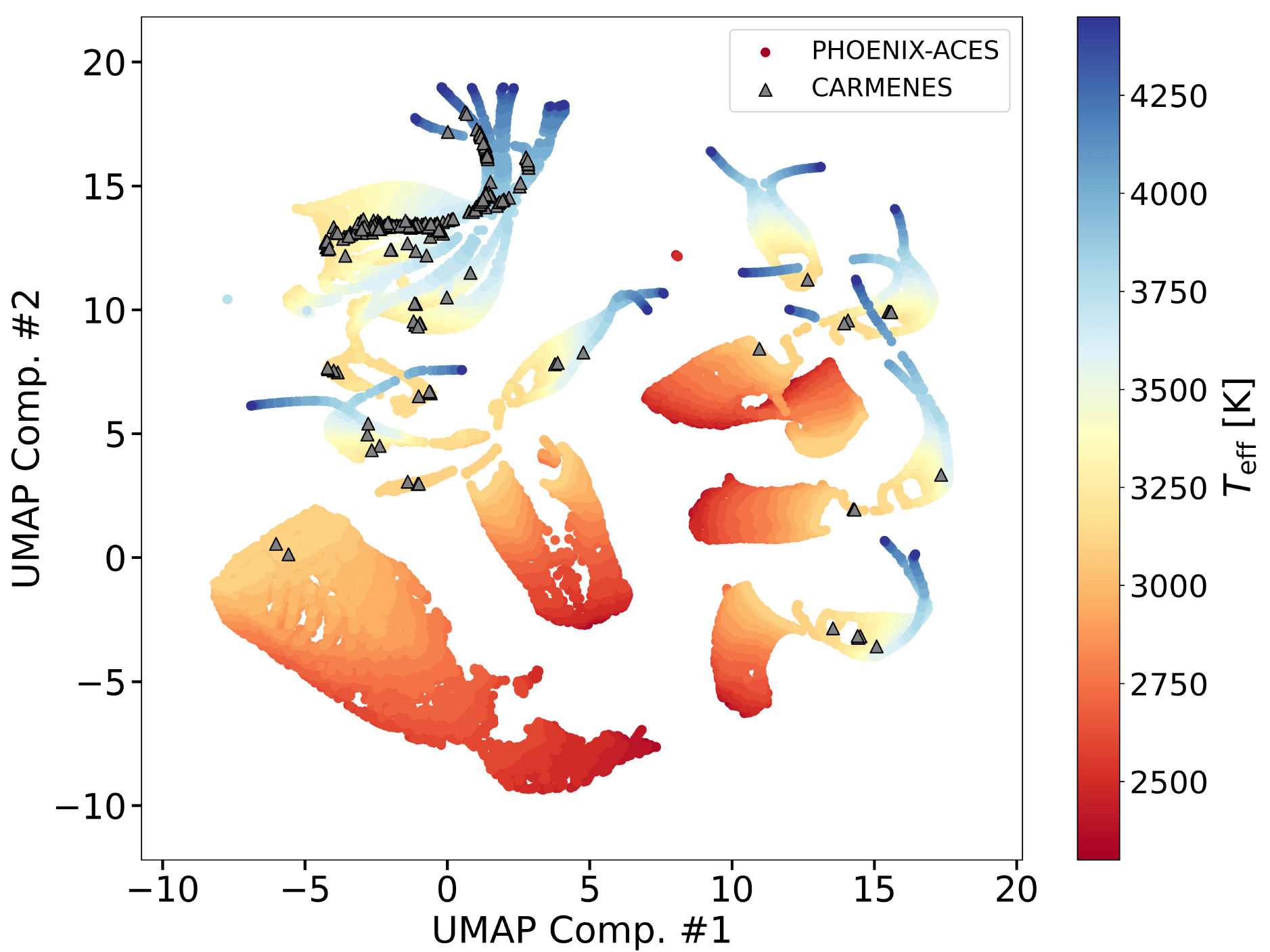}
    \caption{Two-dimensional UMAP projection of one of the 26 sets of PHOENIX-ACES (dots colour-coded by $T_{\rm eff}$) and CARMENES (grey triangles) compressed representations. PHOENIX-ACES encodings are obtained with the initially trained AE and CARMENES encodings with the fine-tuned network.}
    \label{fig:umap_enc}
\end{figure}

While our goal was to preserve the meaningfulness of the low-dimensional representations of the synthetic and observed spectra, we aimed, above all, to minimise the disparity between the observed and synthetic compressed representations. For instance, Fig. \ref{fig:umap_enc} illustrates a UMAP two-dimensional projection, using the same metric as in Fig. \ref{fig:umap_flux}, for one of the 26 sets of PHOENIX-ACES and CARMENES representations. In contrast to Fig. \ref{fig:umap_flux}, in this case, the CARMENES objects are integrated over the space occupied by the PHOENIX-ACES family of projections, leading to a significant reduction of the differences in feature distributions between the two domains. Consequently, we calculated the minimum Euclidean distance from each CARMENES instance to the synthetic grid in both the initial high-dimensional space and the new low-dimensional feature space. While the mean distance is $2.72$ when evaluated in the initial feature space (Fig. \ref{fig:umap_flux}), it is reduced to a mean value of $0.086$ for the encoded representations (Fig. \ref{fig:umap_enc}), averaged over the 26 sets.
In this manner, a latent space that encodes the shared knowledge from both domains was learned, effectively bridging the gap between them.

\subsection{Stellar parameter estimation} \label{sec:cnn}

In the final step of our methodology, we employed CNNs, one of the oldest deep learning approaches \citep{lecun1998}, to estimate the stellar parameters of the 286 CARMENES stars. As a starting point for this process, we used the 26 sets of encoded representations for the PHOENIX-ACES and CARMENES spectra obtained in the previous steps of our work. 

Inspired by the hierarchical structure of the human visual nervous system \citep[a precursor of CNNs; ][]{necognitron1980}, CNNs are therefore generally used to deal with image data. They are a specific class of multilayered feedforward neural networks, initially developed for image classification and visual pattern recognisition \citep{lecun1998,alexnet2012,vggnet2014}. The distinctive factor of CNNs is the use of convolution operations, in the convolutional layers, to automatically extract features from data. After the convolutional structure, the set of features is flattened and passed to an artificial neural network (ANN) to perform the classification or regression task.

In each forward-propagation process, the input of each neuron of the convolutional layer is obtained with an element-wise dot product between a convolution kernel (or filter), with trainable coefficients, and the outputs of the previous layer. The resulting arrays and a tunable bias are added up and passed through an activation function to obtain the output feature map of the neuron. The set of kernels is tuned during the training process, as the weights of the deep ANN layers are adjusted, so that the different feature maps of the layer represent specific features detected in the input data. \citet{surveycnn} provided a detailed review of CNNs.

\begin{figure}
	\includegraphics[width=\columnwidth]{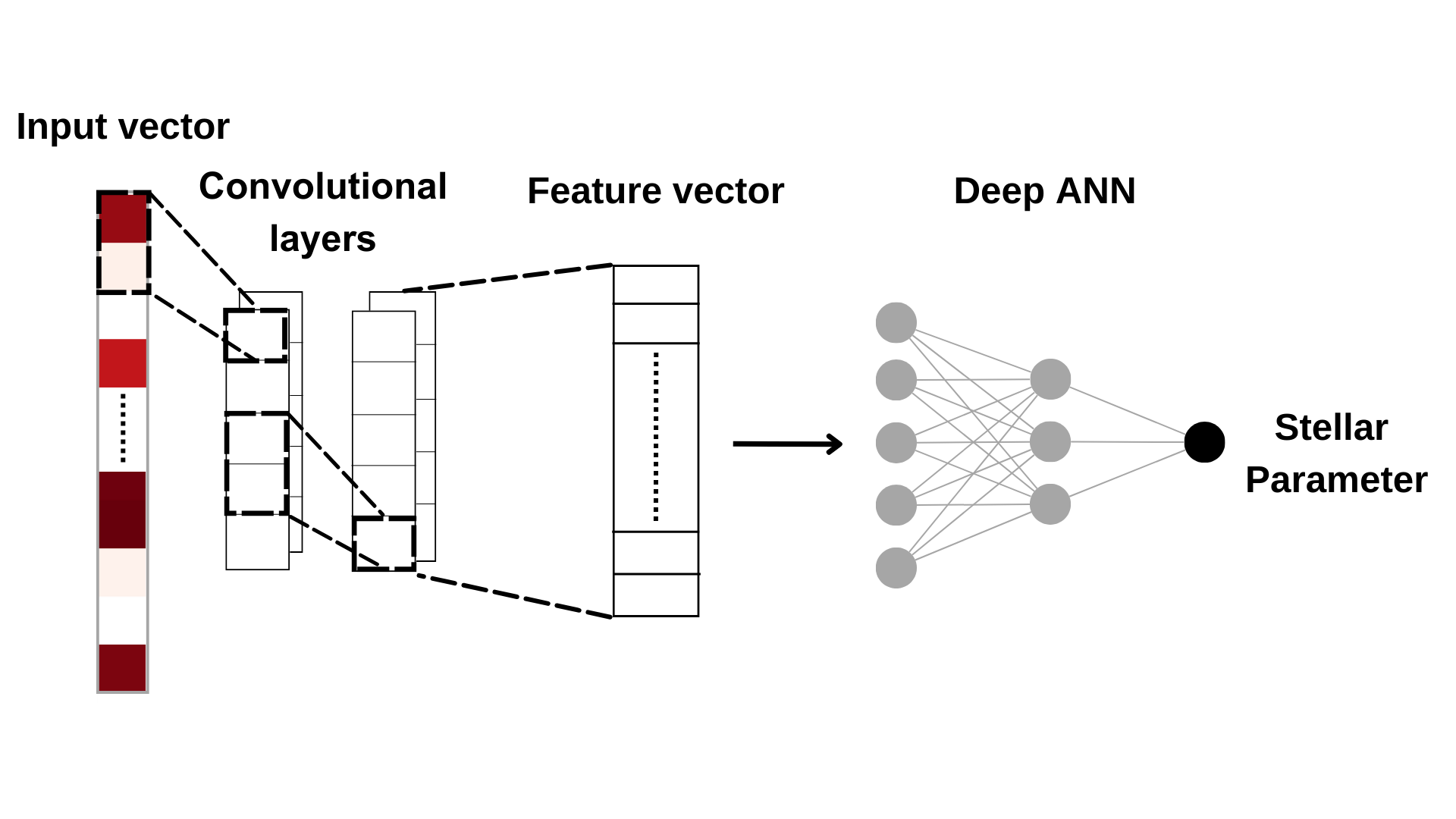}
    \caption{Schematic representation of a one-dimensional CNN architecture.}
    \label{fig:cnn}
\end{figure}

In one-dimensional (1D) CNNs (see Fig. \ref{fig:cnn}), the convolution kernel slides along a sequence of non-independent values to extract relevant features, and they have proven to be highly performant in several applications during the recent years \citep{kiranyaz2019}. \citet{sharma2020} presented a semisupervised learning approach to handle the scarcity of labelled samples, using AE and 1D CNN architectures for stellar spectral classification. \citet{zheng2020} explored how the generation of stellar spectra to balance the training data set can significantly improve the performance of a 1D CNN classifier.

Since we used 32-component vectors as input data for the stellar parameter estimation, we built a 1D CNN architecture. This architecture consists of two convolutional layers (Conv1D) with a variable number of filters (see Table \ref{tab:cnn_arc}), followed by four fully-connected (Dense) layers. A flattening step is incorporated between the convolutional and the ANN components to reshape the output of the final convolutional layer (number of outputs $\times$ number of filters) into a one-dimensional vector. This vector is then fed into the dense layers. We used a rectified linear unit (ReLU) activation function in all layers except the output layer, with a linear activation. We estimated $T_{\rm eff}$, log\,$g$, [M/H], and $v\sin{i}$ independently, searching for the optimal hyperparameters of the 1D CNN architecture (same procedure as in Section \ref{sec:ac}) in the estimation of each parameter. Table \ref{tab:cnn_arc} describes in detail the CNN architectures used. We followed the same procedure in the independent estimation of the different stellar parameters. To have a significant number of final estimates and to assess the robustness of our methodology, we built five CNN models for each of the 26 sets of encoded representations, thus obtaining a total of 130 regressors for each of the parameters.

\begin{table*}
 \caption{CNN architectures used for the estimation of $T_{\rm eff}$, log\,$g$, [M/H], and $v\sin{i}$.}
 \label{tab:cnn_arc}
 \centering
 \begin{tabular}{l c c c c c c c c c c c c}
 
  \hline\hline
  \noalign{\smallskip}

  Layer & \multicolumn{4}{c}{Output size} & \multicolumn{4}{c}{Number of filters} & \multicolumn{4}{c}{Number of parameters} \\

   & $T_{\rm eff}$ & log\,$g$ & [M/H] & $v\sin{i}$ & $T_{\rm eff} $& log\,$g$ & [M/H]  & $v\sin{i}$ & $T_{\rm eff}$ & log\,$g$ & [M/H] & $v\sin{i}$ \\
  
  \noalign{\smallskip}
  \hline
  \noalign{\smallskip}

  Conv1D & 32 & 32 & 32 & 32 & 64 & 16 & 32 & 64 & 192 & 48 & 96 & 192 \\
  Conv1D & 32 & 32 & 32 & 32 & 32 & 64 & 8 & 8 & 4\,128 & 2\,112 & 520 & 1\,032 \\
  Flatten & 1\,024 & 2\,048 & 256 & 256 & \ldots & \ldots & \ldots & \ldots & 0 & 0 & 0 & 0 \\
  Dense & 256 & 256 & 256 & 256 & \ldots & \ldots & \ldots & \ldots & 262\,400 & 524\,544 & 65\,792 & 65\,792 \\
  Dense & 128 & 128 & 128 & 128 & \ldots & \ldots & \ldots & \ldots & 32\,896 & 32\,896 & 32\,896 & 32\,896 \\
  Dense & 64 & 64 & 64 & 64 & \ldots & \ldots & \ldots & \ldots & 8\,256 & 8\,256 & 8\,256 & 8\,256 \\
  Dense & 1 & 1 & 1 & 1 & \ldots & \ldots & \ldots & \ldots & 65 & 65 & 65 & 65 \\
  
  \noalign{\smallskip}
  \hline
 \end{tabular}
\end{table*}

To train the CNN models, we use stratified sampling to create the indices of the traning (70\,\%) and test (30\,\%) sets from the PHOENIX-ACES low-dimensional representations, ensuring that the distribution of the target parameter is representative of the overall distribution in both sets. For this, we relied on the \texttt{StratifiedShuffleSplit} class of the \texttt{scikit-learn} python package, which automatically performs stratification based on a target variable and generates indices to split data into training and test set. We trained the CNN models using the synthetic compressed representations, with a mean squared error loss function, and evaluated them on the test set. As final regressors, we kept the 80 models with the lowest mean squared error in the test set, obtaining an upper value of 353\,K, 0.0042\,dex, 0.0016\,dex, and 0.054\,km\,s$^{-1}$ for $T_{\rm eff}$, log\,$g$, [M/H], and $v\sin{i}$, respectively. Using these models, we obtained 80 final parameter estimates for each of the CARMENES stars.

\begin{figure*}
    \centering
    	\includegraphics[width=\linewidth]{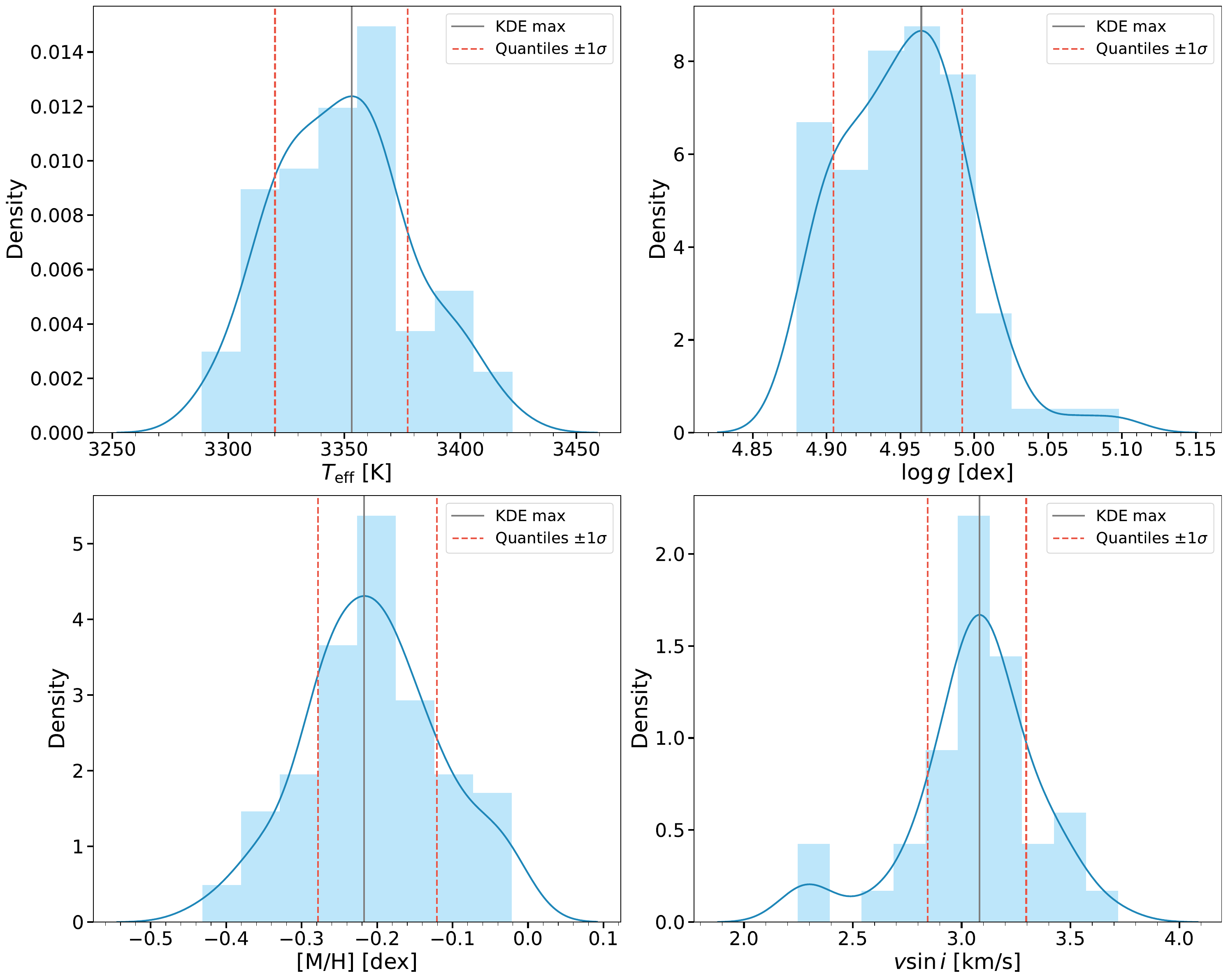}
    \caption{Distribution of stellar parameter estimations of J17578+046 (Barnard's star, M3.5\,V \citep{alonsofloriano2015}). The blue solid line represents the KDE probability density function, with the maximum marked with a grey solid line. The red dashed lines represent the $\pm1\sigma$ uncertainties.}
    \label{fig:kde_plots}
\end{figure*}

We followed the same strategy used by \citetalias{pass20} and \citetalias{bello2023} for the uncertainty estimation of the stellar parameters. For each star, we gathered the 80 estimations and computed the probability density function using the Kernel Density Estimate \citep[KDE; ][]{chen1997, poggio2021} technique. We took the maximum of this probability density function as the confident estimation for the stellar parameter, together with the 1$\sigma$ thresholds as the corresponding uncertainties. Here, the final stellar parameter is derived from a distribution of parameter estimates which come from 26 different sets of input features, together with the five CNN models built for each set. Therefore, the uncertainties provided should be understood as an intrinsic error of our methodology. Fig. \ref{fig:kde_plots} shows an example of the results for a single star.

\section{Results and discussion} \label{sec:results}

\subsection{Stellar parameters analysis} \label{sec:par_analysis}

\begin{figure*}
    \centering
    	\includegraphics[width=8.5cm]{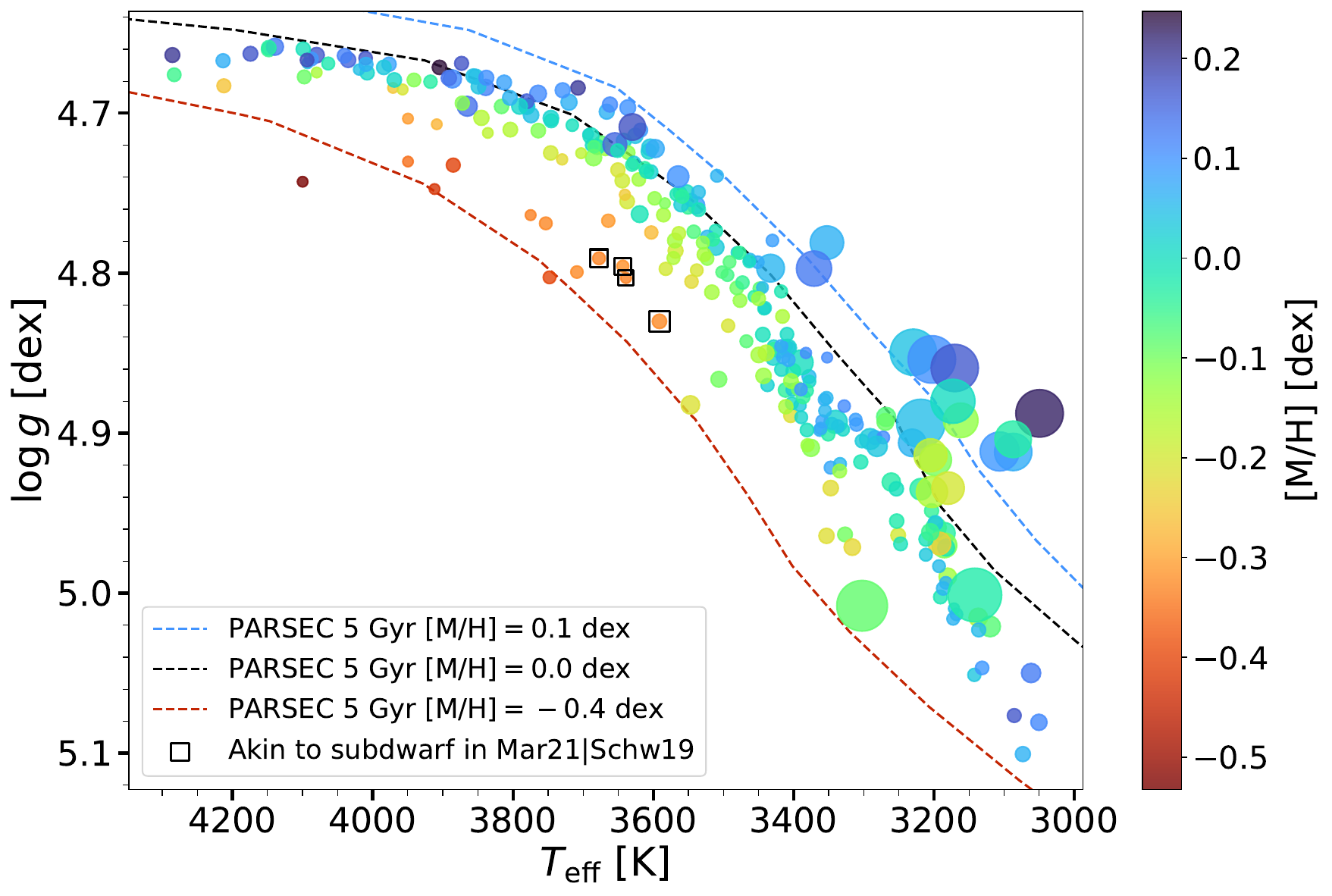}
    	\includegraphics[width=8.5cm]{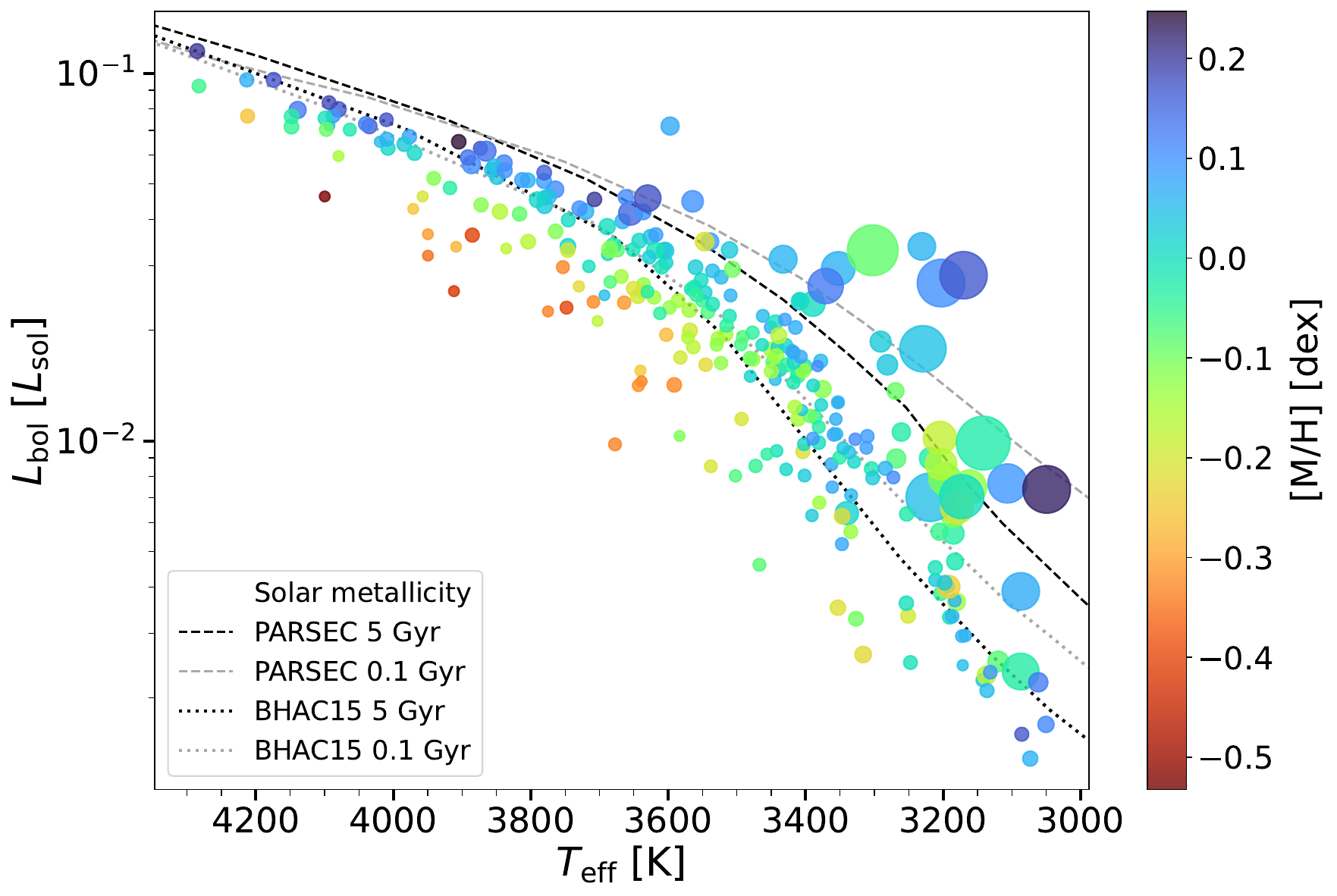}
            \\
    	\includegraphics[width=8.5cm]{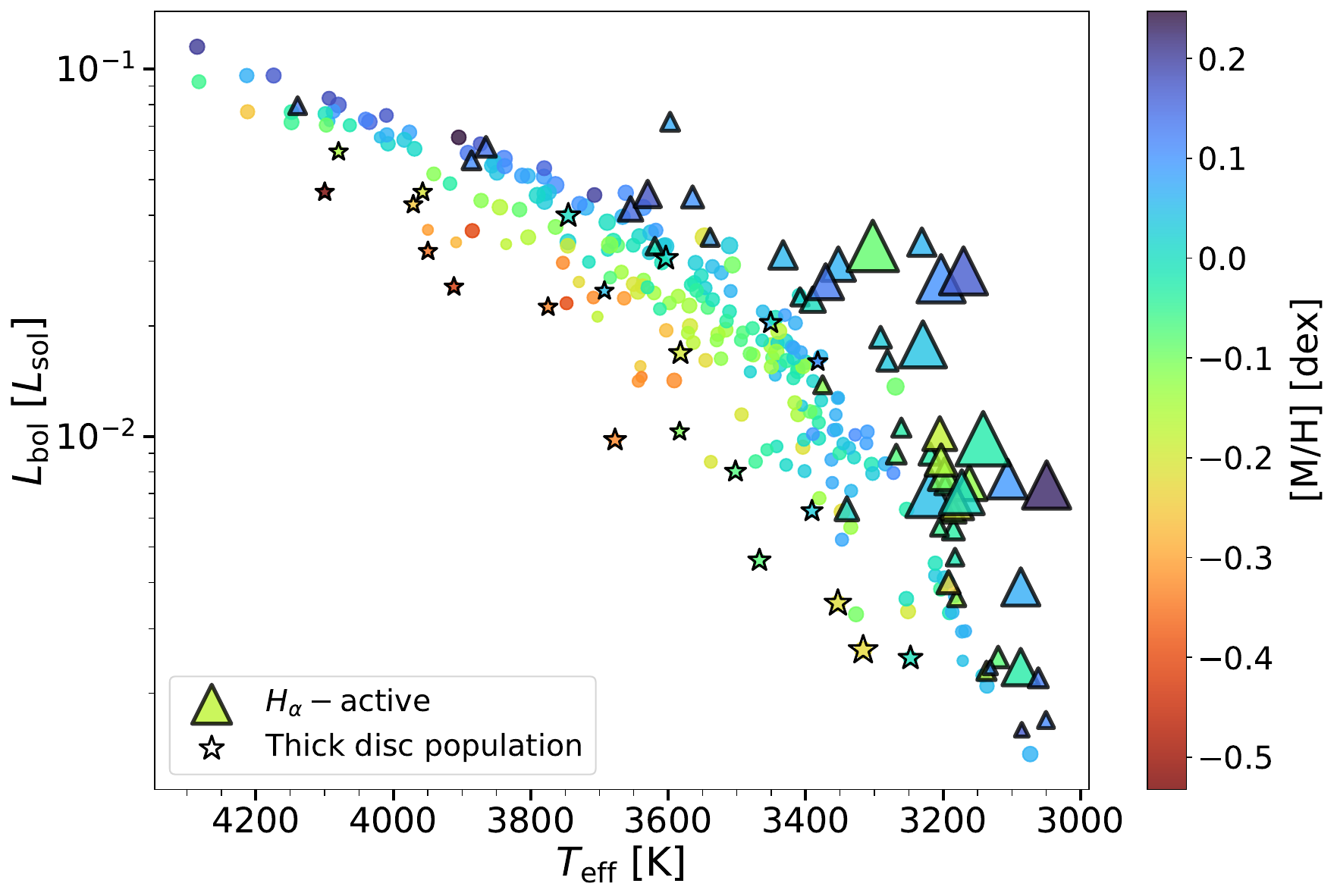}
            \includegraphics[width=8.5cm]{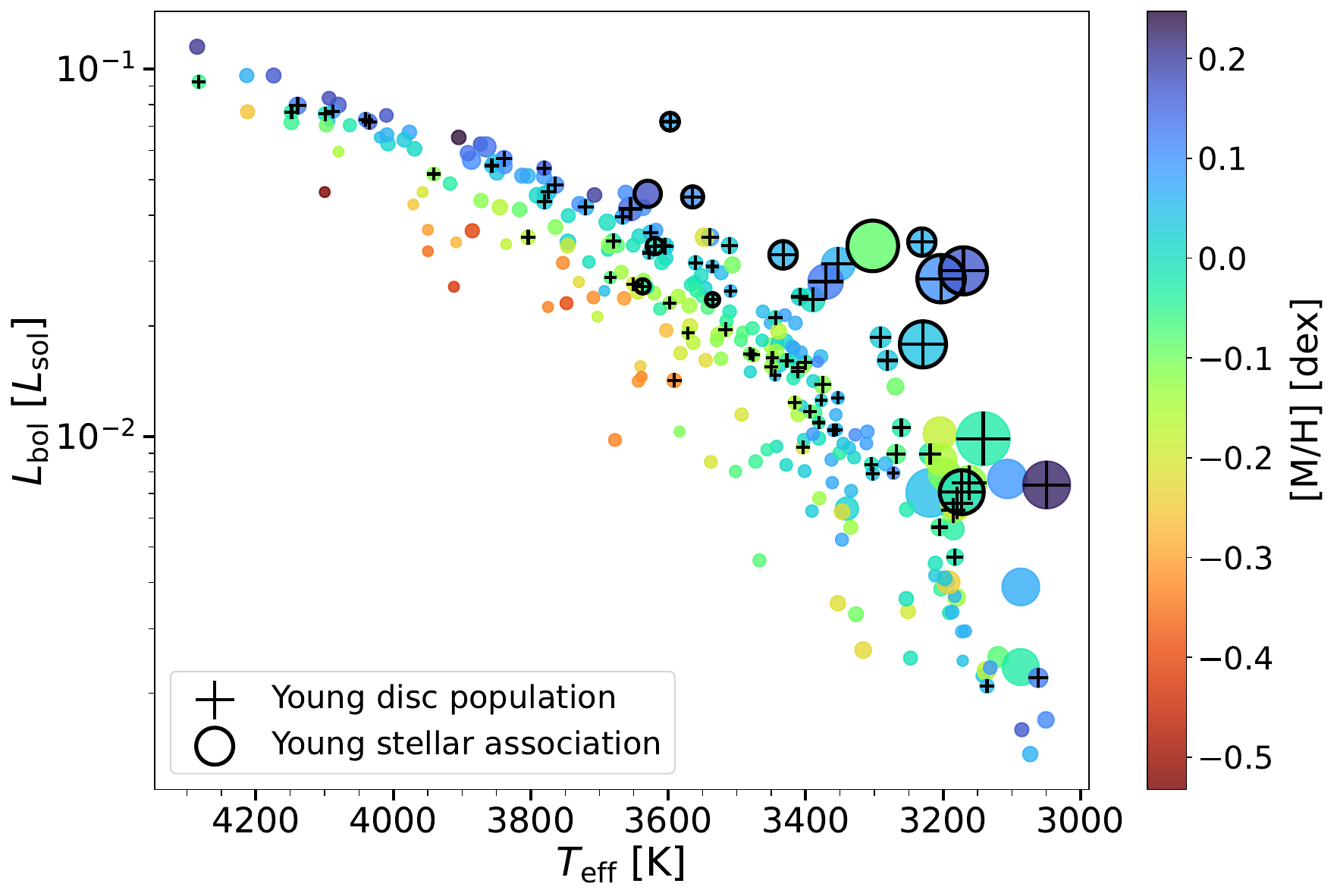}
        \caption{Analysis of the stellar parameters derived with our methodology. The dots are colour-coded according to the estimated metallicity. The size of the dots is proportional to the estimated projected rotational velocity. The \textit{top left panel} shows a Kiel diagram, with the red, black, and blue dashed lines corresponding to 5 Gyr PARSEC isochrones with [M/H] $=-0.4, 0.0$ and $0.1$\,dex, respectively. Empty squares represent the stars reported to have a behaviour akin to subdwarfs both in \citetalias{mar21} and \citetalias{schw19} (same for \textit{bottom left panel}). \textit{Top right:} black and grey dashed lines correspond to solar metallicity PARSEC isochrones for 5 and 0.1 Gyr, respectively. Black and grey dotted lines correspond to solar metallicity \citet{baraffe2015} isochrones for 5 and 0.1 Gyr, respectively. \textit{Bottom left:} triangles represent stars identified as H$\alpha$ active in \citet{schofer2019}. Empty stars depict members of the thick disc Galactic population (Cortés-Contreras et al., in prep.).  \textit{Bottom right:} plus symbols correspond to stars identified as members of the young disc Galactic population by Cortés-Contreras et al. (in prep.). Empty circles represent stars with a possible membership in a young stellar associaton, as explained in Section \ref{sec:par_analysis}.}
    \label{fig:par_diags}
\end{figure*}

Table \ref{tab:pars} presents the stellar atmospheric parameters determined with our methodology. The top left panel in Fig. \ref{fig:par_diags} shows a Kiel diagram that relates all our estimated parameters, along with isochrones based on the  PAdova and TRieste Stellar Evolution Code \citep[PARSEC release v1.2S; ][]{bressan2012} for 5\,Gyr and [M/H] $=-0.4, 0.0,$ and $0.1$\,dex. The results obtained with our methodology follow the trend set by the isochrones and the structure observed in the estimated metallicities is also consistent with them. The remaining three panels in Fig. \ref{fig:par_diags} show a Hertzsprung-Russell diagram (HRD) of our results, with different features highlighted in each of them. We computed the bolometric luminosities, $L_{\mathrm{bol}}$, as \citet{cifuentes2020} using the latest astrometry and photometry from {\it Gaia} DR3 \citep{gaiadr3}. Theoretical isochrones, for solar metallicity, from PARSEC v1.2S and from evolutionary models presented by \citet{baraffe2015} are overplotted in the top right panel for 0.1 and 5\,Gyr. Both the Kiel diagram and the HRD reveal a clear outlier region at the lowest temperatures \citep[mid M-dwarf regime; ][]{cifuentes2020,pecaut2013}, populated mostly by the stars with a high estimated projected rotational velocity ($v\sin{i}$). These fast rotators in our sample are located at the expected M-dwarf regime, following the relation between the spectral types from the CARMENES input catalogue \citep[Carmencita; ][]{alonsofloriano2015,caballero2016a} and the $v\sin{i}$ values calculated by \citet{reiners2018} \citepalias[see Fig. 2 in][]{mar21}.

The bottom panels in Fig. \ref{fig:par_diags} help to understand the outliers that deviate from the main sequence. The bottom left panel shows that almost all the overluminuous outliers in the HRD are identified as H$\alpha$ active stars by \citet{schofer2019}, considered as such if the pseudo-EW of the H$\alpha$ line satisfies pEW$'$(H$\alpha)<-0.3$\,\AA~(H$\alpha$ flag from Table B.1 in \citetalias{mar21}). As found in previous works \citep[e.g. ][]{jeffers2018,reiners2018}, the fraction of H$\alpha$ active stars is higher at later spectral types. There are clear patterns in the HRD which arise from the kinematic membership of the targets. For instance, and in agreement with \citet{jeffers2018}, most H$\alpha$ active and rapidly rotating stars are kinematically young (dots marked with a + in the bottom right panel).

To study the possible membership of our sample to nearby young stellar associatons, we relied on \texttt{BANYAN}~$\Sigma$\footnote{\url{http://www.exoplanetes.umontreal.ca/banyan/}} \citep{banyan}, a Bayesian analysis tool to identify members of young associations. Modelled with multivariate Gaussians in six-dimensional $\rm XYZUVW$ space, \texttt{BANYAN}~$\Sigma$ can derive membership probabilities for all known and well-characterised young associations within 150\,pc. In our case, we used the python version of \texttt{BANYAN}~$\Sigma$\footnote{\url{https://github.com/jgagneastro/banyan_sigma}}, and included the \textit{Gaia} DR3 sky coordinates, proper motion, radial velocity, and parallax of our target stars as input parameters to the algorithm. The classifier gave a high Bayesian probability (>80\,\%) for 9 objects to belong to a young stellar association, in 7 of the cases with a probability greater than 95\,\%. Table \ref{tab:young} lists the details of these objects. All these stars with a possible membership in a young stellar associaton are represented with a thick open circle in the bottom right panel of Fig. \ref{fig:par_diags}. Here, we also considered four extra stars, namely J09133+688 (G\,234-057), J12156+526 (StKM\,2-809), J15218+209 (GJ\,9520), and J18174+483 (TYC\,3529-1437-1), which \citetalias{schw19} mentioned as young age-based outliers.

\begin{table*}
 \caption{Stars in our sample classified by \texttt{BANYAN}~$\Sigma$ with a high Bayesian probability of belonging to a young stellar association.}
 \label{tab:young}
 \centering          
 \begin{tabular}{l l c l l}
  \hline\hline
  \noalign{\smallskip}
  
  Karmn & Name\,$^{(a)}$ & \texttt{BANYAN}~$\Sigma$  Prob.\,$^{(b)}$ & Young association\,$^{(c)}$ & Association reference\\
  
  \noalign{\smallskip}
  \hline
  \noalign{\smallskip}

  J02088+494 & G 173-039 & 99.94\,\% & AB Doradus & \citet{abdmg} \\
  
  \noalign{\smallskip}

  J02519+224 & RBS 365 & 99.79\,\% & $\beta$ Pictoris & \citet{bpictoris} \\
  
  \noalign{\smallskip} 

  J03473-019 & G 080-021 & 99.94\,\% &  AB Doradus & \citet{abdmg} \\
  
  \noalign{\smallskip}
  
  J05019+011\,$^{(d)}$ & 1RXS J050156.7+010845 & 99.91\,\% & $\beta$ Pictoris & \citet{bpictoris} \\

  \noalign{\smallskip} 

  J05062+046\,$^{(d)}$ & RX J0506.2+0439 & 99.79\,\% & $\beta$ Pictoris & \citet{bpictoris} \\
  
  \noalign{\smallskip} 
  
  J09163-186 & LP 787-052 & 95.01\,\% &  Argus & \citet{arg} \\
  
  \noalign{\smallskip}   
  
  J10289+008 & BD+01 2447  & 99.97\,\% &  AB Doradus & \citet{abdmg} \\
  
  \noalign{\smallskip} 
  
  J19511+464 & G 208-042  & 94.17\,\% & Argus & \citet{arg} \\

  \noalign{\smallskip} 
  
  J21164+025 & LSPM J2116+0234  & 85.20\,\% & Argus & \citet{arg} \\  
  
  \noalign{\smallskip}  
  \hline
 \end{tabular}
 \tablefoot{$^{(a)}$ As in \citet{cifuentes2020}. $^{(b)}$ The Bayesian probability that this object belongs to the young stellar association. $^{(c)}$ Most probable Bayesian hypothesis (including the field). $^{(d)}$ Already mentioned in \citetalias{schw19} as candidate members of the corresponding young stellar association.}
\end{table*}

The bottom left panel in Fig. \ref{fig:par_diags} shows that outliers below the main sequence are typically members of the thick disc Galactic population \citep[Cortés-Contreras et al., in prep.; ][]{tesis_miriam}. Furthermore, four of these outliers are reported to have a behaviour akin to subdwarfs (empty squares in top and bottom left panels) both by \citetalias{mar21} and \citetalias{schw19}. Table \ref{tab:subd} details all the outliers we identified with low-metallicity behaviour, along with the metallicity estimations found in the literature. As discussed by \citet{jao2008}, with the decrease in the metallicity of these objects the TiO opacity also strongly decreases, and this less blanketing from the TiO bands causes more continuum flux to radiate from the deeper and hotter layer of the stellar atmosphere, so that these stars appear bluer than their solar metallicity counterparts (see Fig. 1 in \citet{jao2008}). Our [M/H] determinations for these stars are, in general, in good agreement with the literature.

\begin{table*}
 \caption{Low-metallicity stars identified in Fig. \ref{fig:par_diags}.}
 \label{tab:subd}
 \centering          
 \begin{tabular}{l l c c c c c c}
  \hline\hline
  \noalign{\smallskip}
  
  Karmn & Name & [Fe/H]$_{\mathrm{AE}}^{\,(a)}$ & [Fe/H]$_{\mathrm{DTL}}^{\,(b)}$ & [Fe/H]$_{\mathrm{Mann15}}^{\,(c)}$ & [Fe/H]$_{\mathrm{corr, Mar21}}^{\,(d)}$ & [Fe/H]$_{\mathrm{Schw19}}^{\,(e)}$ & Pop.$^{\,(f)}$\\

    &  & [dex] & [dex]  & [dex] & [dex] & [dex] & \\
  
  \noalign{\smallskip}
  \hline
  \noalign{\smallskip}
  
  J00183+440 & GX And & $-0.33_{-0.17}^{+0.06}$ & $-0.26_{-\ldots}^{+\ldots}$ & $-0.30\pm0.08$ & $-0.52\pm0.11$ & $-0.25\pm0.16$ & D \\
  
  \noalign{\smallskip}
  
  J02123+035 & BD+02 348 & $-0.35_{-0.10}^{+0.12}$ & $-0.33_{-0.01}^{+0.01}$ & $-0.36\pm0.08$ & $-0.49\pm0.06$ & $-0.05\pm0.16$ & TD \\

  \noalign{\smallskip}
  
  J06371+175 & HD 260655 & $-0.41_{-0.13}^{+0.11}$ & $-0.37_{-0.02}^{+0.02}$ & $-0.34\pm0.08$ & $-0.43\pm0.04$ & $-0.42\pm0.16$ & TD-D \\

  \noalign{\smallskip}

  J11033+359 & Lalande 21185 & $-0.34_{-0.13}^{+0.08}$ & $-0.31_{-\ldots}^{+\ldots}$ & $-0.38\pm0.08$ & $-0.49\pm0.10$ & $-0.09\pm0.16$ & TD\\

  \noalign{\smallskip}
  
  J11054+435 & BD+44 2051A & $-0.40_{-0.18}^{+0.07}$ & $-0.35_{-\ldots}^{+\ldots}$ & $-0.37\pm0.08$ & $-0.56\pm0.09$ & $-0.3\pm0.16$ & TD-D \\

  \noalign{\smallskip}

  J12248-182\,$^{(g)}$ & Ross 695 & $-0.33_{-0.18}^{+0.06}$ & $-0.40_{-0.04}^{+0.02}$ & \ldots & $-0.60\pm0.09$ & $-0.17\pm0.16$ & TD \\

  \noalign{\smallskip}

  J13450+176 & BD+18 2776 & $-0.53_{-0.27}^{+0.09}$ & $-0.46_{-0.05}^{+0.06}$ & $-0.54\pm0.08$ & $-0.54\pm0.03$ & $-0.43\pm0.16$ & TD \\

  \noalign{\smallskip}

  J16254+543\,$^{(g)}$ & GJ 625 & $-0.33_{-0.15}^{+0.05}$ & $-0.32_{-0.03}^{+0.02}$ & $-0.35\pm0.08$ & $-0.28\pm0.07$ & $-0.26\pm0.16$ & YD\\

  \noalign{\smallskip}

  J17378+185 & BD+18 3421 & $-0.33_{-0.08}^{+0.11}$ & $-0.23_{-0.03}^{+0.01}$ & $-0.25\pm0.08$ & $-0.40\pm0.07$ & $-0.23\pm0.16$ & D \\

  \noalign{\smallskip}
  
  J19070+208\,$^{(g)}$ & Ross 730 & $-0.34_{-0.18}^{+0.05}$ & $-0.32_{-0.02}^{+0.01}$ & $-0.33\pm0.08$ & $-0.46\pm0.07$ & $-0.20\pm0.16$ & D \\

  \noalign{\smallskip}
  
  J19072+208\,$^{(g)}$ & HD 349726 & $-0.32_{-0.17}^{+0.05}$ & $-0.32_{-0.02}^{+0.02}$ & $-0.35\pm0.08$ & $-0.46\pm0.06$ & $-0.23\pm0.16$ & D \\

  \noalign{\smallskip}
  
  J23492+024 & BR Psc & $-0.43_{-0.12}^{+0.11}$ & $-0.40_{-0.03}^{+0.02}$ & $-0.45\pm0.08$ & $-0.55\pm0.08$ & $-0.13\pm0.16$ & TD \\

  \noalign{\smallskip}
  \hline
 \end{tabular}
 \tablefoot{As explained in \citet{pass20,passegger2022}, our [M/H] results directly translate into [Fe/H] values. $^{(a)}$ From this work. $^{(b)}$ From \citetalias{bello2023}. $^{(c)}$ From \citet{mann2015}. $^{(d)}$ From \citetalias{mar21}, corrected from the $\alpha$ enhancement. $^{(e)}$ From \citetalias{schw19}. $^{(f)}$ Galactic populations, including the thick disc (TD), the thick disc-thin disc transition (TD-D), the thin disc (D), and the young disc (YD), following Cortés-Contreras et al. in prep. $^{(g)}$ Reported to have a behaviour akin to subdwarfs in \citetalias{mar21} or \citetalias{schw19}. In particular, J19070+208 and J19072+208 are both components of the wide binary system LDS\,1017, and \citet{houdebine2008} already identified them as subdwarfs.}
\end{table*}

\begin{figure}
	\includegraphics[width=\columnwidth]{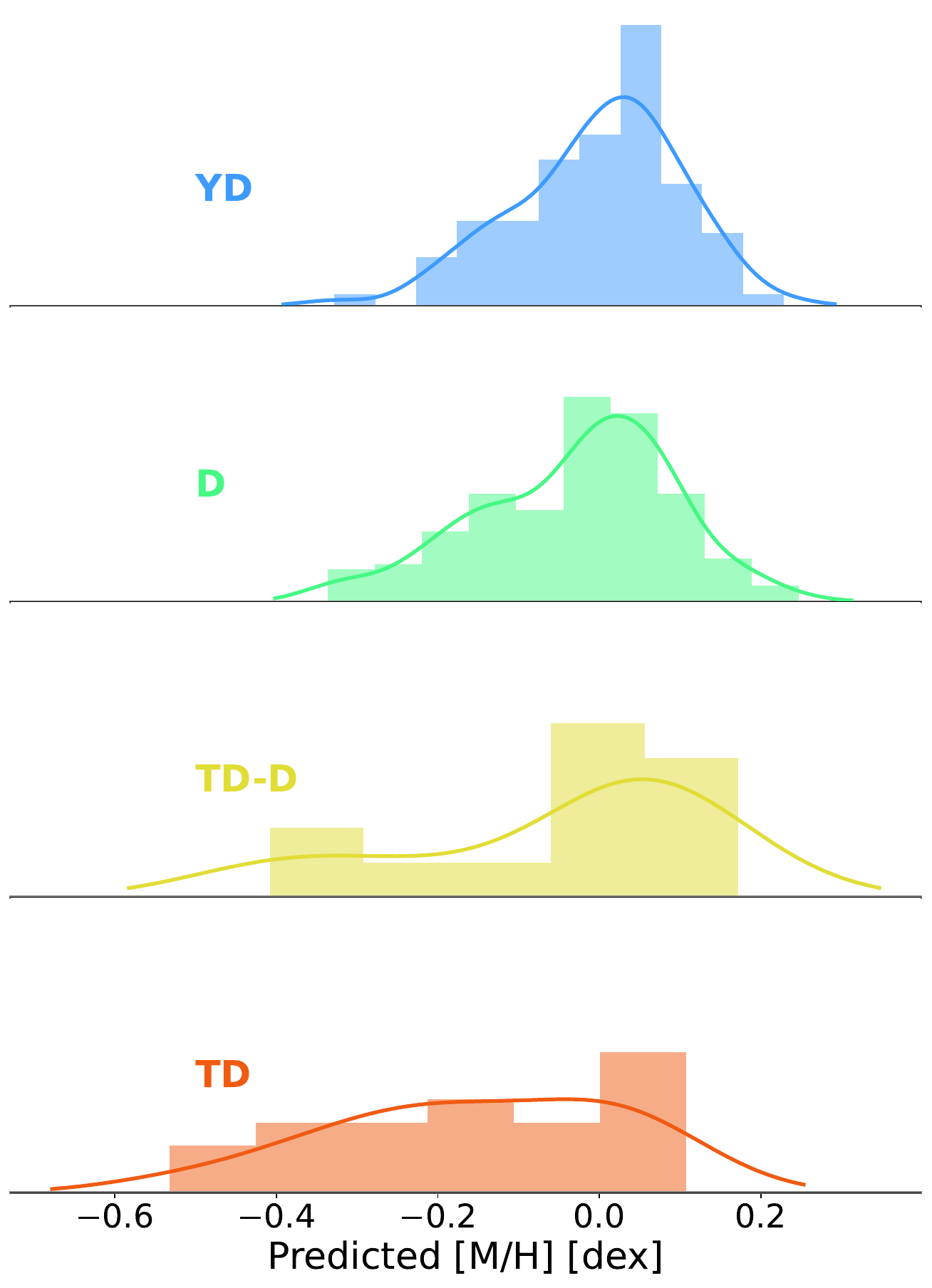}
    \caption{Distribution of our predicted metallicities broken down by kinematic membership in the the thick disc (TD), the thick disc-thin disc transition (TD-D), the thin disc (D), and the young disc (YD) Galactic populations \citep[Cortés-Contreras et al., in prep.; ][]{tesis_miriam}. The bins are normalised so that the total area of the histogram equals one,} and the solid lines represent the KDE probability density function.
    \label{fig:pop_mh_hist}
\end{figure}

Fig. \ref{fig:pop_mh_hist} shows the distribution of our predicted metallicities broken down by kinematic membership in the thick disc (TD), thick disc-thin disc transition (TD-D), thin disc (D), and young disc (YD) Galactic populations \citep[Cortés-Contreras et al., in prep.; ][]{tesis_miriam}. This breakdown reveals the distinction between metal-rich thin disc stars and metal-poor stars in the older thick disc \citep{bensby2005,gaiamh2023}, with the TD-D transition as an intermediate step. To prove this, we performed a two-sample Kolmogorov-Smirnov test \citep{kolmogorov33,smirnov48} on the thin and thick disc samples, which returned a $p\,\rm{value} = 0.0071$, rejecting the hypothesis that both samples come from the same distribution.

\begin{figure}
	\includegraphics[width=\columnwidth]{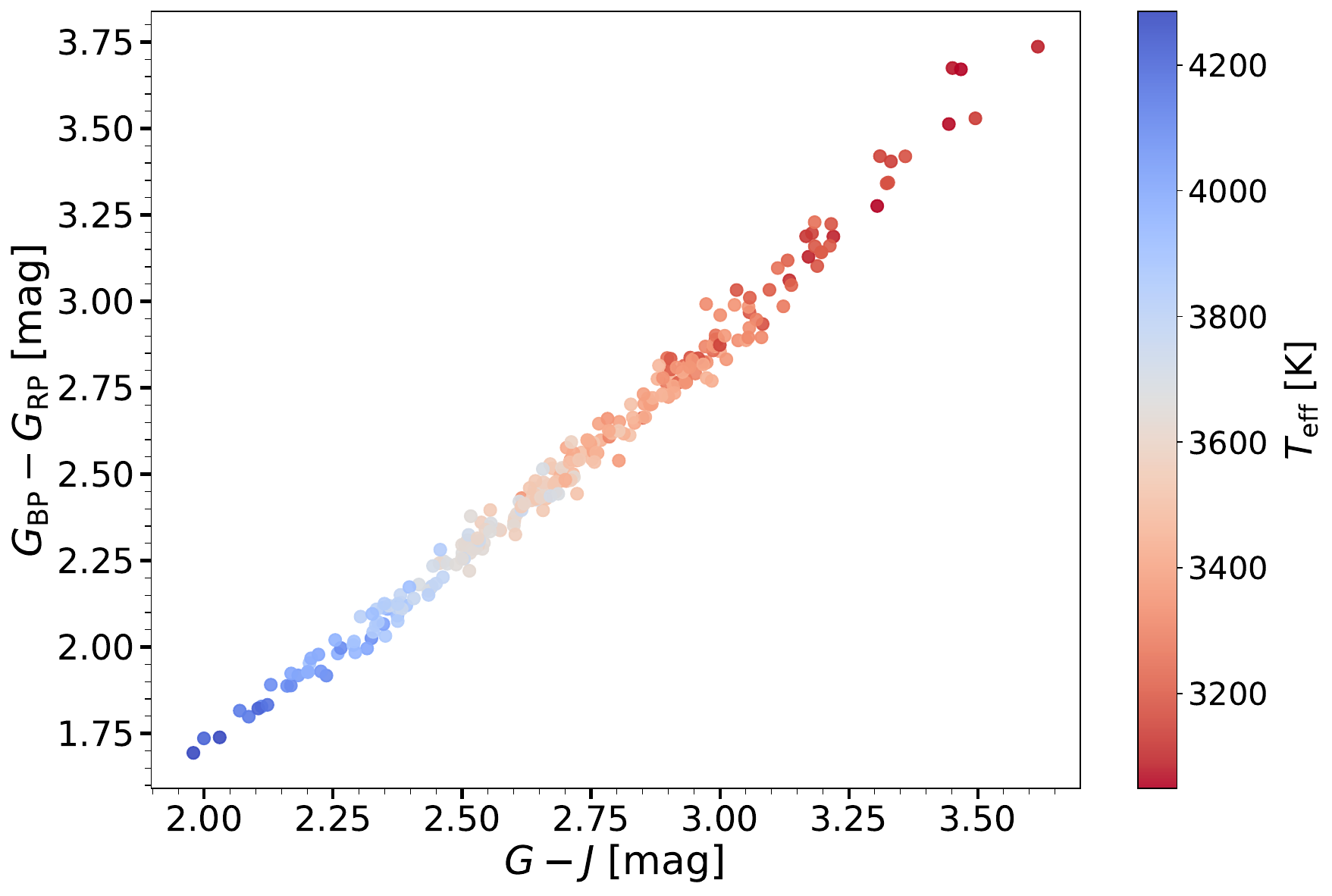}
    \caption{2MASS-\textit{Gaia} G$_{\mathrm{BP}} - G_{\mathrm{RP}}$ versus $G-J$ diagram of our target stars with good photometric quality (2MASS \texttt{Qflg}=A and a relative error of less than 10\,\% in \textit{Gaia}\,DR3 photometry). The points are colour-coded according to the effective temperatures derived in this work.}
    \label{fig:col_diag}
\end{figure}

Also, the 2MASS-\textit{Gaia} G$_{\mathrm{BP}} - G_{\mathrm{RP}}$ versus $G-J$ colour-colour diagram in Fig. \ref{fig:col_diag} shows how the evolution in our estimated effective temperatures is coherent with the colour-colour relationship (see Fig. 14 in \citealt{cifuentes2020}). For this diagram, we only considered stars with reliable 2MASS $J$-band and \textit{Gaia} DR3 G$_{\mathrm{BP}}$ and G$_{\mathrm{RP}}$ photometry.

\subsection{Comparison with the literature}
\label{sec:lit_comp}

We compared our results with different collections found in the literature. Whereas this section focuses on the latest studies using CARMENES data, namely \citetalias{bello2023}, \citetalias{mar21}, \citetalias{pass2019}, \citetalias{pass20}, and \citetalias{schw19}, a more extensive compilation of literature, together with the uncertainties of the estimations, is provided in Appendix \ref{app_a}. For \citetalias{pass2019}, we considered the parameters derived from VIS spectra. Table \ref{tab:comparison_lit} lists the mean difference ($\overline{\Delta}$; literature$-$this work), root mean squared error (rmse), and Pearson correlation coefficient ($r_{\rm p}$) for the comparison with each of the literature collections. An interactive version of the results presented in this section is available to the astronomical community\,\footnote{\url{https://cab.inta-csic.es/users/pmas/}}.

\begin{table*}
 \caption{Mean difference ($\overline{\Delta}$; literature$-$this work), root mean square error (rmse), and Pearson correlation coefficient ($r_{\rm p}$) for the comparison between our results and the literature.}
 \label{tab:comparison_lit}
 \centering
 \begin{tabular}{l c c c c c c c c c c c c}
 
  \hline\hline
  \noalign{\smallskip}

  Reference & \multicolumn{3}{c}{$T_{\rm eff}$\,[K]} & \multicolumn{3}{c}{log\,$g$\,[dex]} & \multicolumn{3}{c}{[Fe/H]\,[dex]} & \multicolumn{3}{c}{$v\sin{i}$\,[km\,s$^{-1}$]} \\

     & $\overline{\Delta}$ & rmse & $r_{\rm p}$ & $\overline{\Delta}$ & rmse & $r_{\rm p}$ & $\overline{\Delta}$ & rmse & $r_{\rm p}$ & $\overline{\Delta}$ & rmse & $r_{\rm p}$ \\
  
  \noalign{\smallskip}
  \hline
  \noalign{\smallskip}

  \citetalias{bello2023} & -117 & 180 & 0.87 & \ldots & \ldots & \ldots & 0.01 & 0.14 & 0.60 & \ldots & \ldots & \ldots \\
  \citetalias{mar21} & -19 & 102 & 0.94 & 0.12 & 0.18 & 0.39 & -0.11 & 0.16 & 0.65 & \ldots & \ldots & \ldots \\
  \citetalias{pass2019} & -80 & 117 & 0.96 & 0.00 & 0.05 & 0.86 & 0.06 & 0.15 & 0.52 & \ldots & \ldots & \ldots \\
  \citetalias{pass20} & -35 & 51 & 0.99 & -0.04 & 0.06 & 0.93 & 0.23 & 0.25 & 0.76 & 1.64 & 1.94 & 0.99 \\
  Rein18$^{\,(a)}$ & \ldots & \ldots & \ldots & \ldots & \ldots & \ldots & \ldots & \ldots & \ldots & -0.86 & 1.51 & 0.98 \\
  \citetalias{schw19} & -40 & 93 & 0.96 & 0.13 & 0.14 & 0.89 & 0.00 & 0.10 & 0.63 & \ldots & \ldots & \ldots \\
  
  \noalign{\smallskip}
  \hline
 \end{tabular}
 \tablefoot{$^{(a)}$ From \citet{reiners2018}.}
\end{table*}

\begin{figure*}
    \centering
    	\includegraphics[width=8.5cm]{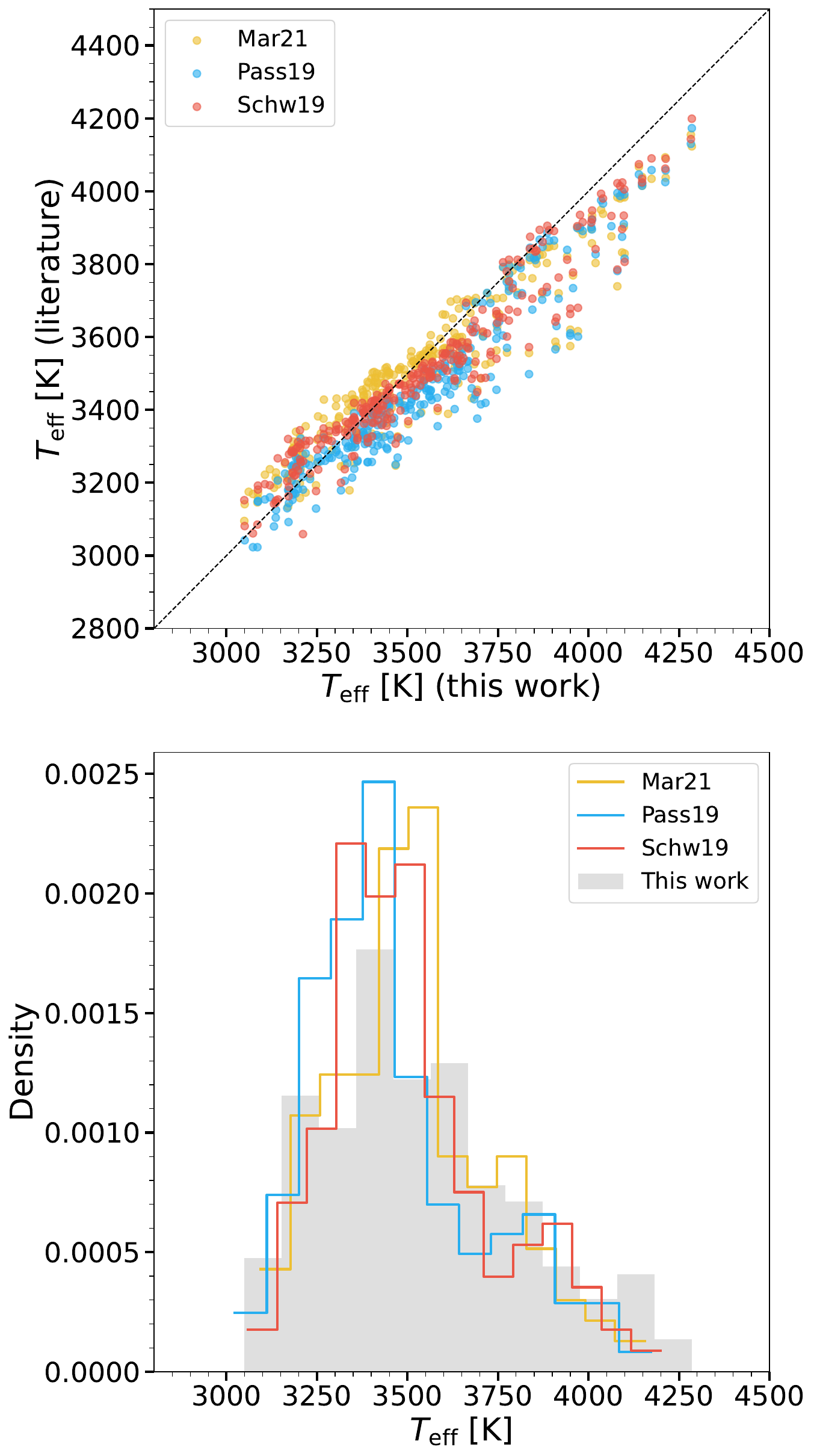}
            \includegraphics[width=8.5cm]{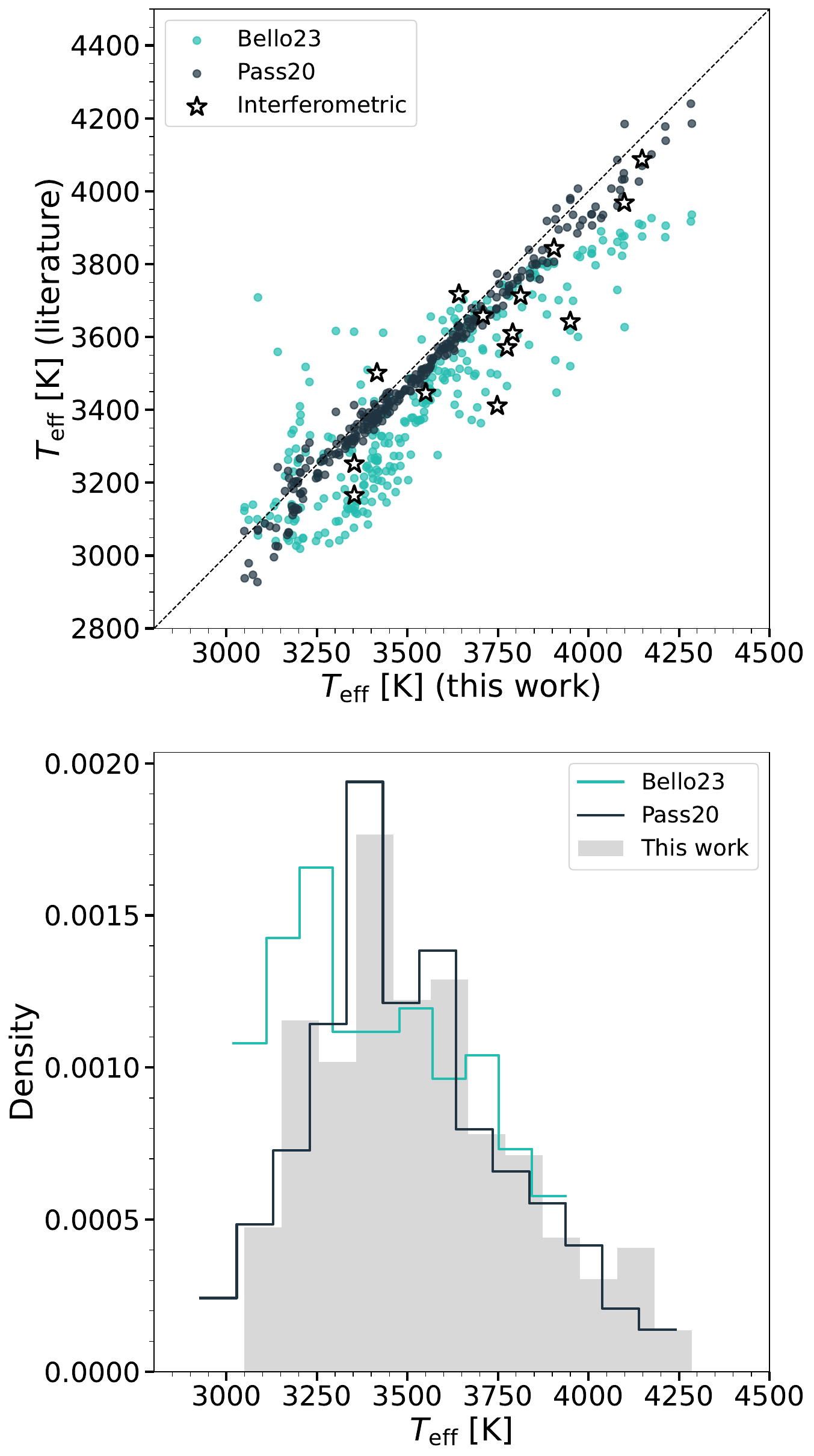}
    \caption{Comparison between our derived $T_{\rm eff}$ values and the literature. The \textit{left panels} include the results from \citetalias{mar21} (yellow), \citetalias{pass2019} (blue), and \citetalias{schw19} (red). The \textit{right panels} include the work from \citetalias{bello2023} (cyan) and the results obtained following the DL methodology described by \citetalias{pass20} (dark blue). The black stars in the \textit{top right panel} correspond to the interferometrically derived $T_{\rm eff}$ values from \citetalias{bello2023}. The dashed black lines in the \textit{top panels} correspond to the 1:1 relation. For the bin width in the histograms shown in the \textit{bottom panels}, we used the default parameters of the \texttt{seaborn histplot} function.}
    \label{fig:scatter_teff}
\end{figure*}

Figure~\ref{fig:scatter_teff} depicts the comparison with literature values for $T_{\rm eff}$. The left panels show a similar linear trend among \citetalias{mar21}, \citetalias{pass2019}, and \citetalias{schw19} with our values, all of them with a slope of less than one, for the region $T_{\rm eff}$ (this work) $\lesssim 3\,750$\,K. From this value onwards, where the number of stars in our training set is smaller, the dispersion increases significantly and our $T_{\rm eff}$ estimations deviate towards hotter values, resulting in a mean difference of $\overline{\Delta}=-19$\,K, $\overline{\Delta}=-80$\,K, $\overline{\Delta}=-40$\,K for \citetalias{mar21}, \citetalias{pass2019}, and \citetalias{schw19}, respectively. The figures provided in Appendix \ref{app_a} show that the uncertainties intrinsic to our methodology are also larger for estimations above 3\,750\,K. The right panels show how the agreement with the values obtained following the approach described by \citetalias{pass20} is excellent, which is expected since their methodology is the closest to the one presented in this work. Moreover, the comparison with the results from \citetalias{bello2023} reveals the same structure, but inverted, as shown in Fig. 9 of their work, with a larger dispersion than that observed for the other literature collections. The black stars in the top right panel represent the 14 interferometrically derived $T_{\rm eff}$ values (see Table 1 in \citetalias{bello2023}), which are on average cooler than the temperatures obtained with our methodology ($\overline{\Delta}_{\rm interf}=-119$\,K). The $r_{\rm p}$ values listed in Table~\ref{tab:comparison_lit} show a strong correlation with all the collections.

\begin{figure*}
    \centering
    	\includegraphics[width=8.5cm]{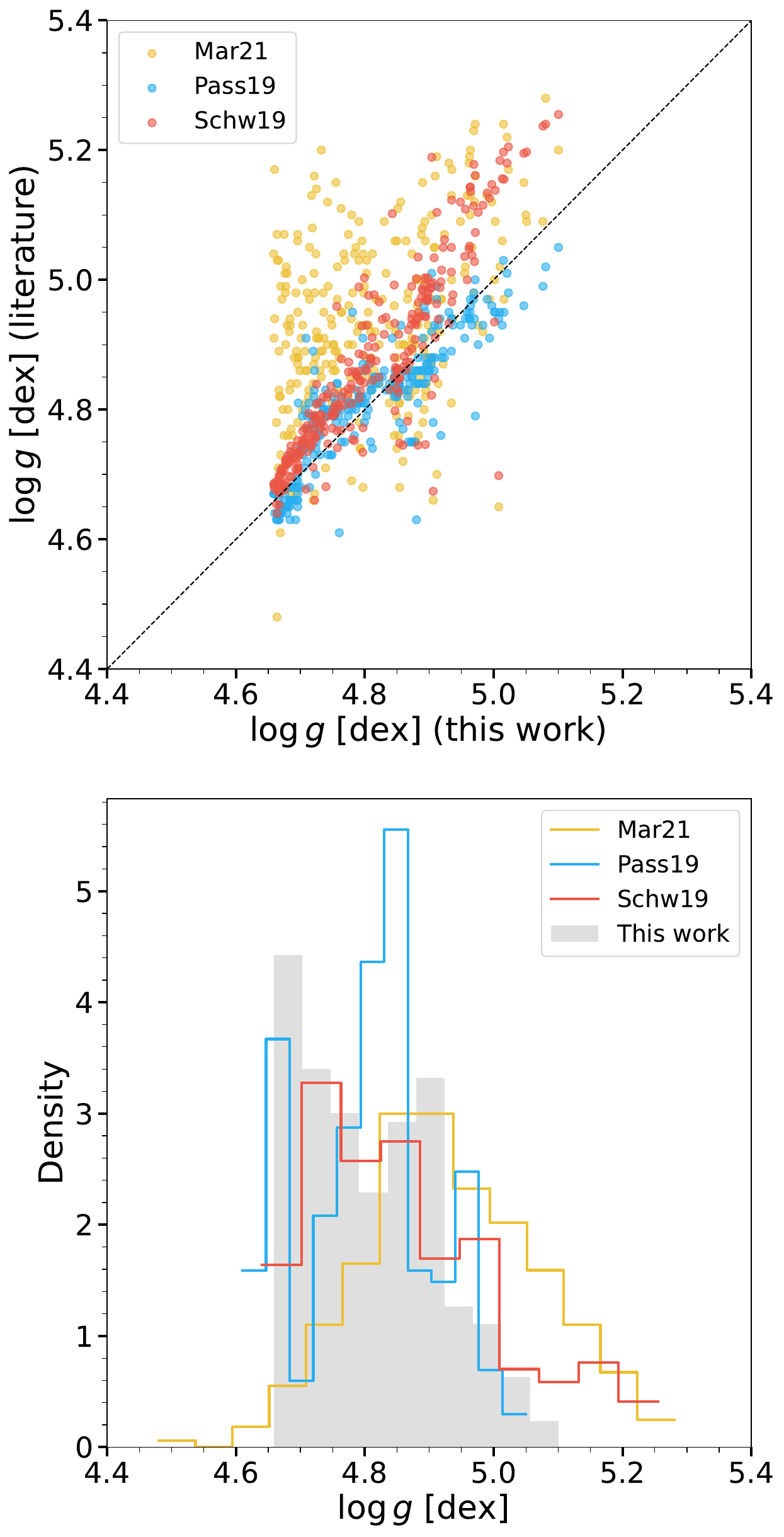}
            \includegraphics[width=8.5cm]{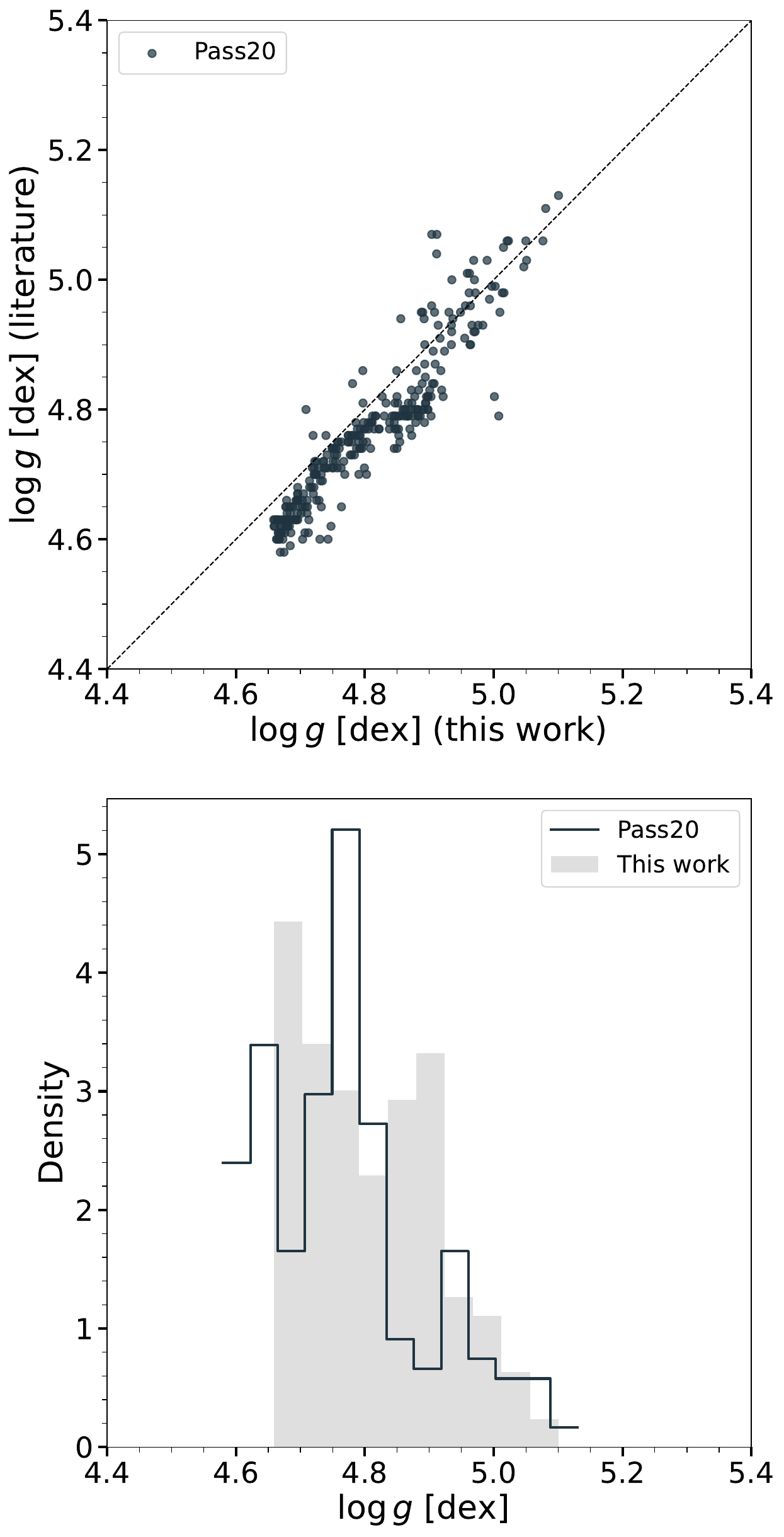}
    \caption{Comparison between our derived log\,$g$ values and the literature. Colours and symbols are the same as in Fig. \ref{fig:scatter_teff}.}
    \label{fig:scatter_logg}
\end{figure*}

Figure \ref{fig:scatter_logg} shows a similar literature comparison for log\,$g$. For \citetalias{schw19}, we considered the values derived using their mass-radius relation and the Stefan-Boltzmann's law. The log\,$g$ values from \citetalias{mar21} show a large dispersion ($r_{\rm p}=0.39$), as already mentioned in their work, and are generally spread towards higher values ($\overline{\Delta}=0.12$\,dex). While the results from \citetalias{pass2019} cover the same range and are similar on average to our obtained log\,$g$ ($\overline{\Delta}=0.00$\,dex), those from \citetalias{schw19} extend to higher values and are on average higher than ours ($\overline{\Delta}=0.13$\,dex). It should be noted that, while \citetalias{pass2019} and \citetalias{schw19} fix log\,$g$ using theoretical isochrones, \citetalias{mar21} has log\,$g$ as a free parameter.
Moreover, our results show a good correlation ($r_{\rm p}=0.93$) with those obtained following the methodology described by \citetalias{pass20}, although the latter are deviated to lower values ($\overline{\Delta}=-0.04$\,dex).

\begin{figure*}
    \centering
    	\includegraphics[width=8.5cm]{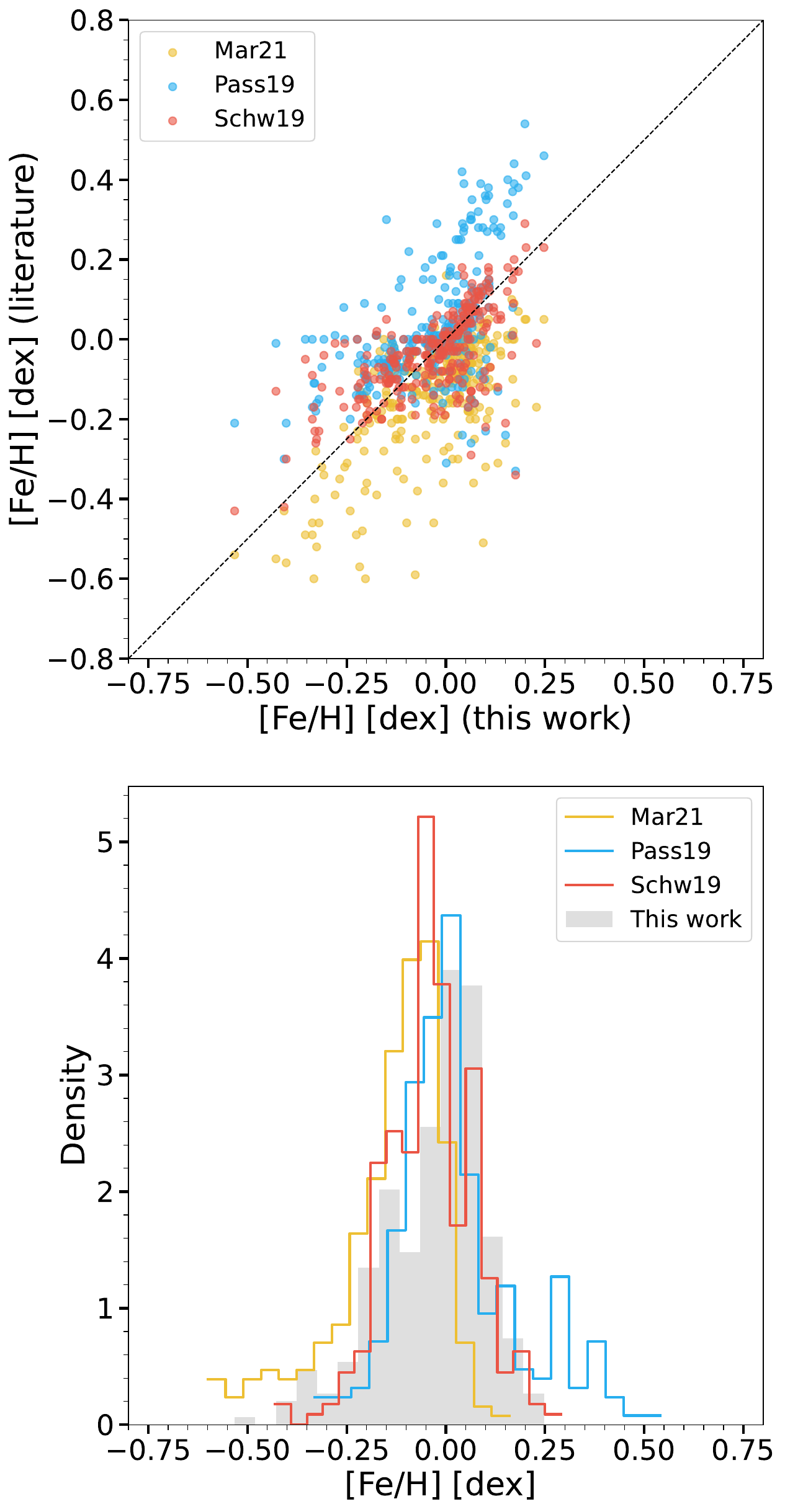}
            \includegraphics[width=8.5cm]{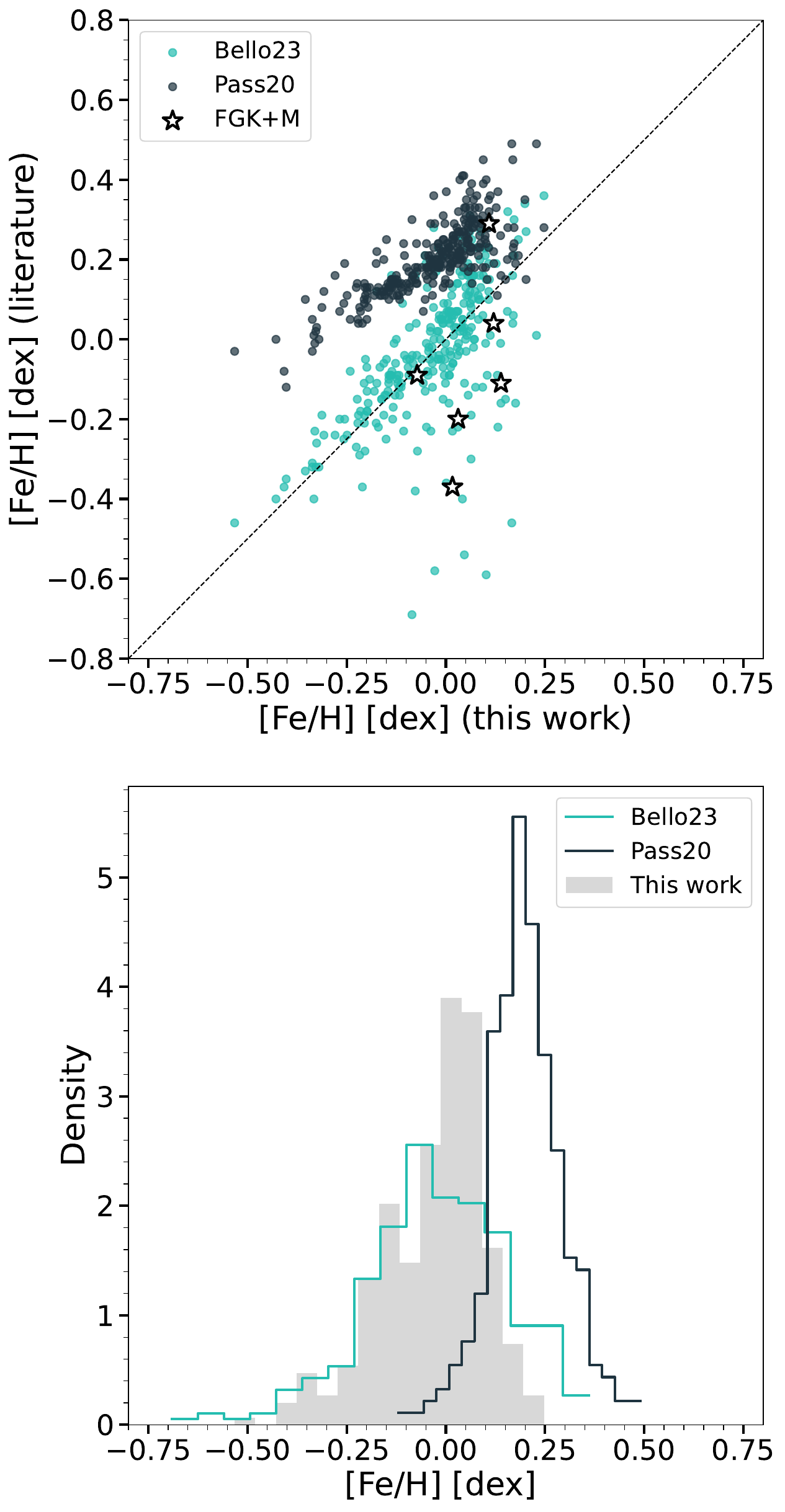}
    \caption{Comparison between our derived [Fe/H] values and the literature. Colours and symbols are the same as in Fig. \ref{fig:scatter_teff}. The black stars in the \textit{top right panel} correspond to the spectroscopically determined [Fe/H] values from FGK+M systems presented in \citetalias{bello2023}.}
    \label{fig:scatter_mh}
\end{figure*}

As discussed in \citet{passegger2022}, several discrepancies can be found when comparing metallicities of M dwarfs obtained with different methodologies. Figure \ref{fig:scatter_mh} shows the comparison with literature values for our [M/H] estimations, which directly translate into [Fe/H] values \citep{pass20,passegger2022}. For \citetalias{mar21}, we considered the values corrected for alpha enhancement. Our results are similar on average to those from \citetalias{schw19} ($\overline{\Delta}=0.00$\,dex), while \citetalias{pass2019} and \citetalias{mar21} results tend to be higher and lower, with $\overline{\Delta}=0.06$ and $\overline{\Delta}=-0.11$\,dex, respectively. As already mentioned in \citet{passegger2022}, the results from the DL methodology described by \citetalias{pass20} are deviated towards more metal-rich values, with $\overline{\Delta}=0.23$\,dex. We note that this deviation, which is attributed to the synthetic gap by \citetalias{pass20}, does not appear in the DTL methodologies presented by \citetalias{bello2023} and here. \citetalias{bello2023} metallicities cover more or less the same range as our results, and the spectroscopically determined [M/H] values from FGK+M systems (see Table 3 in \citetalias{bello2023}) (black stars in the top right panel) are systematically lower ($\overline{\Delta}=-0.13$\,dex).

\begin{figure}
	\includegraphics[width=\columnwidth]{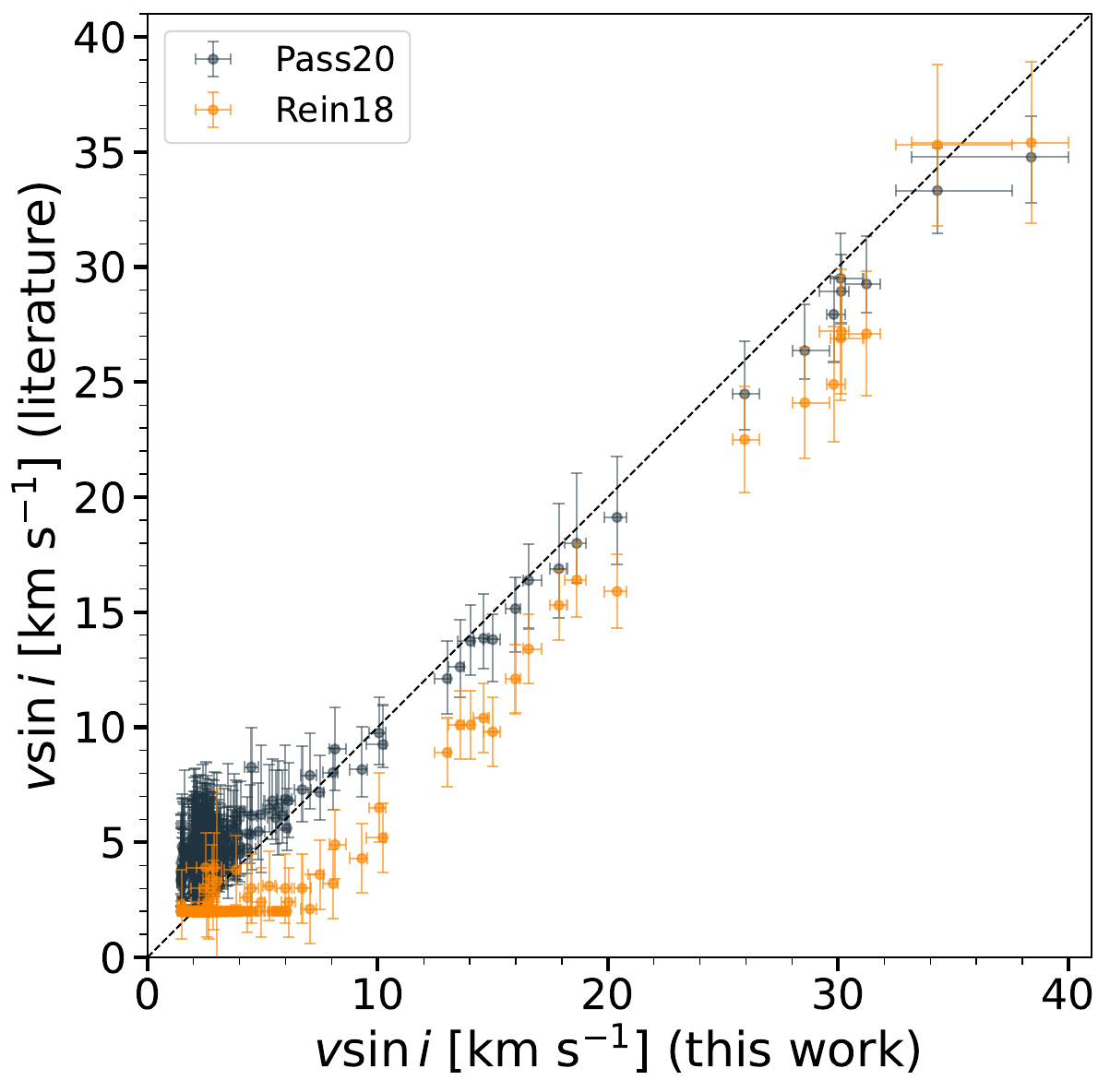}
    \caption{Comparison between our derived $v\sin{i}$ values and the literature. The `Rein18' label stands for the results presented in \citet{reiners2018}.}
    \label{fig:scatter_vsini}
\end{figure}

We also compared our $v\sin{i}$ determinations with the ones derived by \citet{reiners2018} using the cross-correlation method and with those obtained following the DL methodology described by \citetalias{pass20}. Fig. \ref{fig:scatter_vsini} shows how our derived $v\sin{i}$ are mostly consistent with the literature within their errors. Both \citetalias{pass20} and \citet{reiners2018} results show a good correlation with our values ($r_{\rm p}=0.99$ and 0.98, respectively). Since most of the objects are located at lower $v\sin{i}$ values, it is convenient to split the analysis provided in Table \ref{tab:comparison_lit} at a cut-off value of $v\sin{i}\,{\rm (this\,work)}=12$\,km\,s$^{-1}$. Below this value, \citetalias{pass20} presents $\overline{\Delta}=1.83$\,km\,s$^{-1}$ and ${\rm rmse}=1.97$\,km\,s$^{-1}$, with $\overline{\Delta}=-1.22$\,km\,s$^{-1}$ and ${\rm rmse}=1.45$\,km\,s$^{-1}$ for faster rotators. Similarly, for \citet{reiners2018}, we obtained $\overline{\Delta}=-0.68$\,km\,s$^{-1}$ and ${\rm rmse}=1.24$\,km\,s$^{-1}$ for values below the threshold, and $\overline{\Delta}=-3.47$\,km\,s$^{-1}$ and ${\rm rmse}=3.71$\,km\,s$^{-1}$ for values above.


\section{Conclusions}\label{sec:conclusions}

This work serves as an extension of a series of papers (\citetalias{pass20}; \citetalias{bello2023}) dedicated to exploring the use of DL for stellar parameter estimation of CARMENES M dwarfs, based on synthetic spectra. \citetalias{bello2023} developed a model-based DTL technique to bridge the significant differences in flux features between the two spectral families, reported by \citetalias{pass20}. Here, we propose a parallel feature-based DTL strategy that addresses the limitations mentioned in their work regarding the need for high-quality stellar parameter estimations in the knowledge transfer process. All the resources, including the code developed to build the methodology described in Section \ref{sec:methodology} and the code to reproduce the figures displayed in Section \ref{sec:results} are publicly available at \texttt{GitHub}\footnote{\url{https://github.com/pedromasb/autoencoders-CARMENES}}.

Using a methodology that combines the use of AEs and CNNs, we derived new estimations for the stellar parameters $T_{\rm eff}$, log\,$g$, [M/H], and $v\sin{i}$ of 286 M dwarfs observed with CARMENES. The AE models were trained on PHOENIX-ACES synthetic spectra and then fine-tuned using the CARMENES high-S/N, high-resolution spectra. In the fine-tuning process, no data other than the observed spectra are required, which gives our methodology great flexibility, as no measured stellar parameters are involved in the knowledge transfer. We used the low-dimensional representations of the synthetic and observed spectra, resulting from the initial training and the fine-tuning steps, respectively, as input to the CNNs for the estimation of the stellar parameters. In this way, parameter estimation is conducted using a dataset in which no significant differences in the feature distributions between the synthetic and observed data are evident.

We performed an in-depth analysis of our estimated stellar parameters, using the diagram shown in Fig. \ref{fig:par_diags} to study the objects that deviate from the main sequence. We found that almost all the overlumimuous outliers are identified as H$\alpha$ active stars by \citet{schofer2019}, while outliers located below the main sequence are typically metal-poor stars from the thick disc Galactic population. In particular, using the \texttt{BANYAN}~$\Sigma$ tool, we found 9 objects with a high Bayesian probability of belonging to five different young stellar associations, in 7 of these cases with a probability of more than 95\,\%. Together with the low-metallicity objects already reported in \citetalias{mar21} and \citetalias{schw19}, we identified eight more stars that exhibit the same behaviour.

We also conducted a comparative study between our results and the latest studies using CARMENES data, finding good consistency with the literature in most cases. Both our $T_{\rm eff}$ and log\,$g$ determinations are, in general, strongly correlated with the results from the literature, with a systematic deviation in our $T_{\rm eff}$ scale towards hotter values for estimations above 3\,750\,K. As expected, our parameter determinations are in very good agreement with \citetalias{pass20}, since their methodology is the most similar to the one presented in this paper. More importantly, the deviation in metallicity attributed to the synthetic gap in their work is not observed in ours thanks to the DTL approach. This, together with the work presented by \citetalias{bello2023}, demonstrates the great potential of DTL-based strategies to bridge the synthetic gap in stellar parameter estimation from synthetic spectra.


\begin{acknowledgements}

We thank the anonymous referee for the comments that helped to improve the quality of this paper. We acknowledge financial support from the Agencia Estatal de Investigaci\'on (AEI/10.13039/501100011033) of the Ministerio de Ciencia e Innovaci\'on and the ERDF `A way of making Europe' through projects PID2022-137241NB-C4[2,4],	  
PID2020-112949GB-I00 (Spanish Virtual Observatory \url{https://svo.cab.inta-csic.es}),
PID2020-117493GB-I00,         
PID2019-109522GB-C5[1,4],     
and grant PR47/21 TAU-CM PRTR-CM,
the Instituto Nacional de T\'ecnica Aeroespacial through grant PRE-OVE, 
and the Gobierno de Canarias through project ProID2020010129. 

We made extensive use of Python throughout the entire process, including the packages \texttt{pandas}\footnote{\url{https://github.com/pandas-dev/pandas}}, \texttt{seaborn} \citep{seaborn}, \texttt{numpy} \citep{numpy}, \texttt{matplotlib} \citep{matplotlib}, \texttt{scikit-learn}  \citep{sklearn}, \texttt{tensorflow} \citep{tensorflow}, \texttt{plotly}\footnote{\url{https://github.com/plotly/plotly.py}}, \texttt{scipy} \citep{scipy}, and \texttt{umap-learn} \citep{mcinnes2018umap}.

\end{acknowledgements}

\bibliographystyle{aa} 
\bibliography{Bibliography} 

\appendix


\section{Additional tables}
\label{app_b}

Table \ref{tab:pars} is available in its entirety in electronic form at the CDS. This appendix only shows an extract of the table to facilitate its understanding.

\begin{table*}
 \caption{Stellar atmospheric parameters, together with their uncertainties, determined with our methodology. Only the first five rows of the table are shown.}
 \label{tab:pars}
 \centering          
 \begin{tabular}{l l c c c c c c}
  \hline\hline
  \noalign{\smallskip}
  
  Karmn & Name & $\alpha\,^{(a)}$ & $\delta\,^{(a)}$ & $T_{\rm eff}$ & log\,$g$ & [Fe/H] & $v\sin{i}$ \\

    & & [J2016.0] & [J2016.0] & [K]  & [dex] & [dex] & [km\,s$^{-1}$] \\
  
  \noalign{\smallskip}
  \hline
  \noalign{\smallskip}
  
  J00051+457 & GJ 2 & 00:05:12.22 & 03:03:08.6 & $3780_{-34}^{+41}$ & $4.70_{-0.04}^{+0.01}$ & $0.03_{-0.04}^{+0.06}$ & $3.19_{-0.16}^{+0.48}$ \\
  
  \noalign{\smallskip}
  
  J00067-075 & GJ 1002 & 00:06:42.32 & 23:29:48.8 & $3073_{-27}^{+18}$ & $5.10_{-0.09}^{+0.04}$ & $0.06_{-0.15}^{+0.08}$ & $3.02_{-0.66}^{+0.33}$ \\
  
  \noalign{\smallskip}

  J00162+198E & LP 404-062 & 00:16:16.96 & 01:19:26.6 & $3362_{-25}^{+34}$ & $4.90_{-0.06}^{+0.02}$ & $0.07_{-0.17}^{+0.02}$ & $2.13_{-0.22}^{+0.30}$ \\
  
  \noalign{\smallskip}

  J00183+440 & GX And & 00:18:27.17 & 02:56:05.9 & $3709_{-43}^{+15}$ & $4.80_{-0.07}^{+0.03}$ & $-0.33_{-0.17}^{+0.06}$ & $2.02_{-0.30}^{+0.20}$ \\
  
  \noalign{\smallskip}

  J00184+440 & GQ And & 00:18:30.07 & 02:56:06.9 & $3251_{-13}^{+36}$ & $4.96_{-0.03}^{+0.05}$ & $-0.20_{-0.10}^{+0.09}$ & $2.82_{-0.24}^{+0.22}$ \\
  
  \noalign{\smallskip}
  \hline
 \end{tabular}
 \tablefoot{$^{(a)}$  From {\it Gaia} DR3.}
\end{table*}


\section{Additional comparison with the literature} \label{app_a}

\begin{figure*}
    \centering
   	\includegraphics[width=0.68\linewidth]{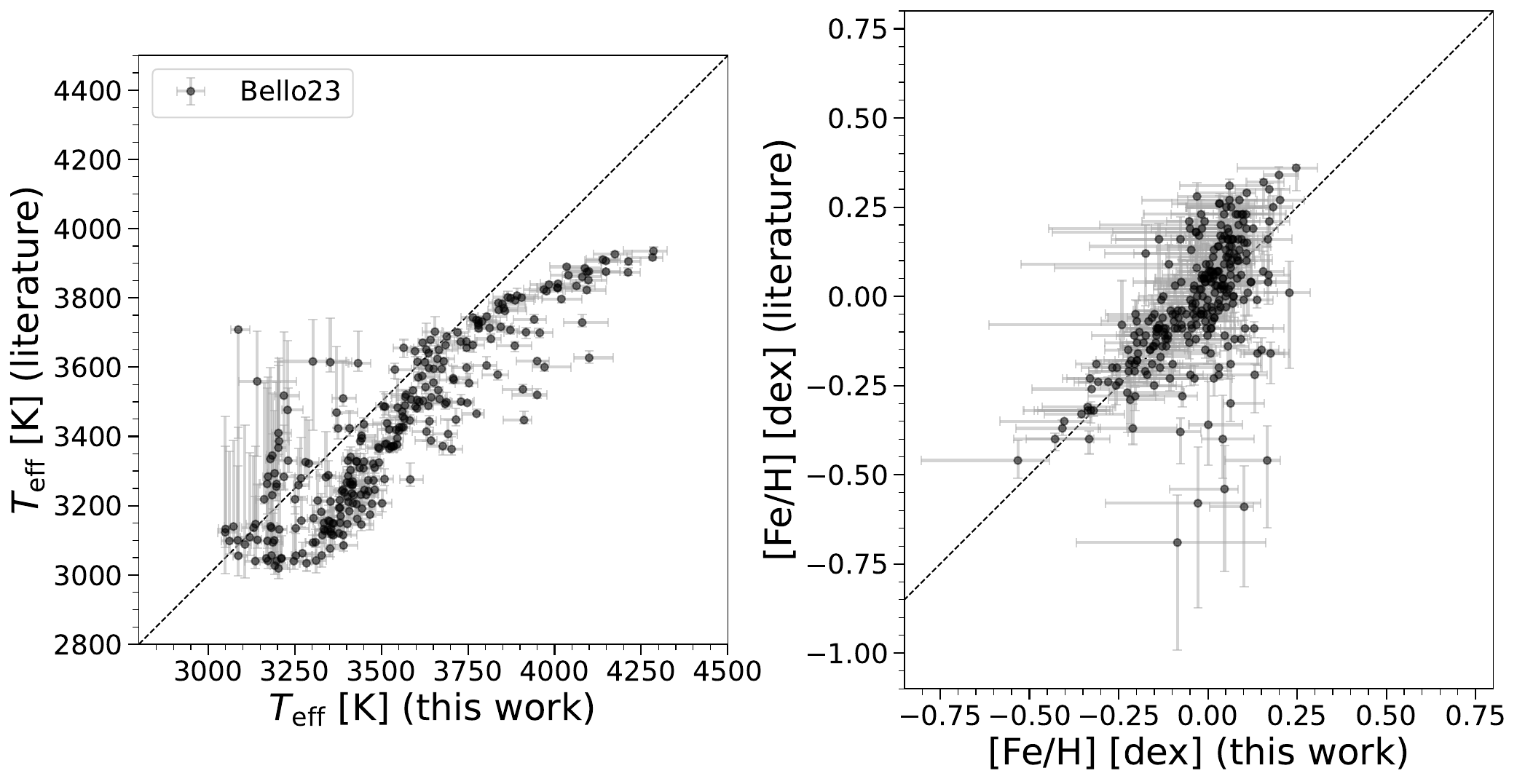}
    \caption{Comparison with \citetalias{bello2023}.}
    \label{fig:appendix_bello23}
\end{figure*}

\begin{figure*}
    \centering
    	\includegraphics[width=\linewidth]{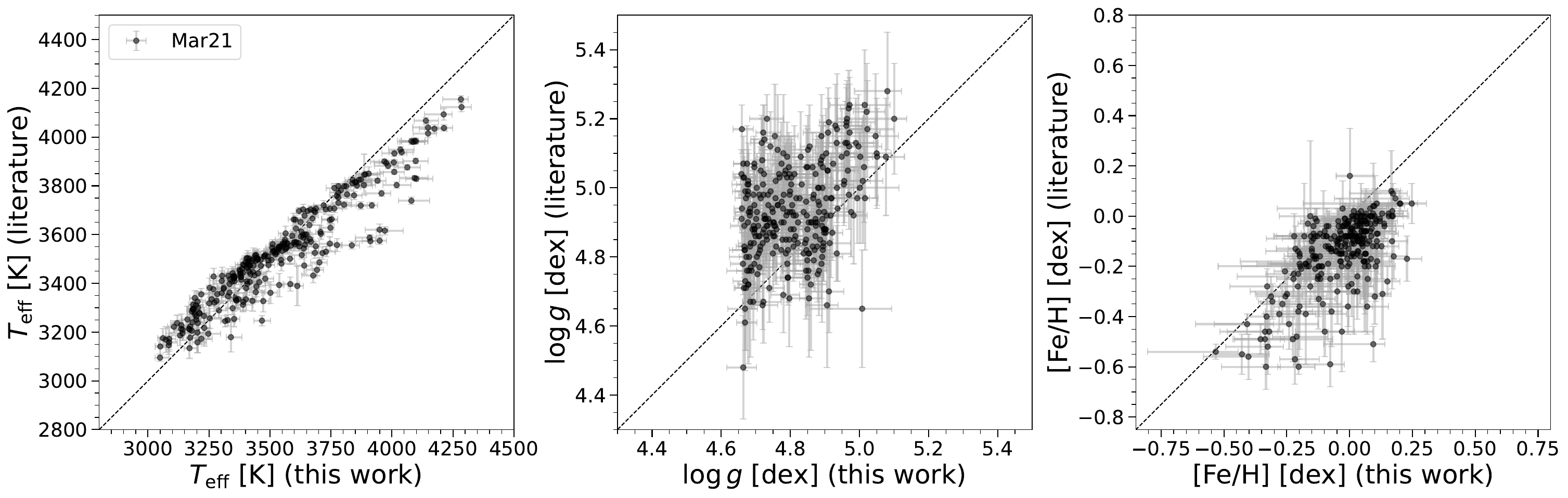}
    \caption{Comparison with \citetalias{mar21}.}
    \label{fig:appendix_mar21}
\end{figure*}

\begin{figure*}
    \centering
    	\includegraphics[width=\linewidth]{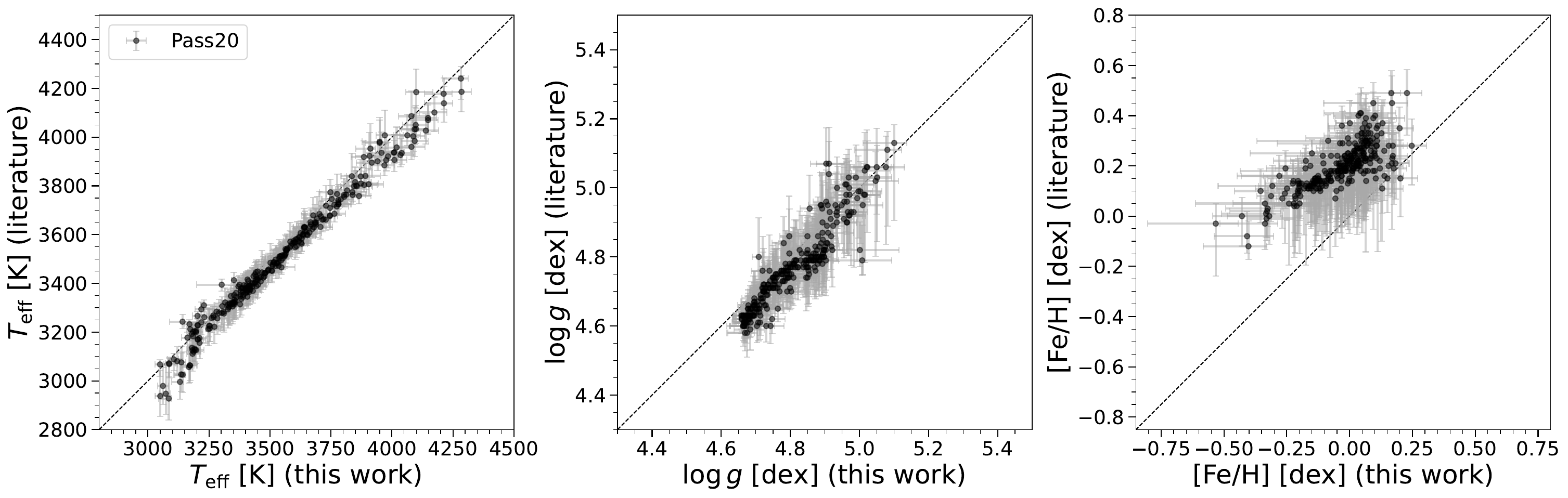}
    \caption{Comparison with \citetalias{pass20}.}
    \label{fig:appendix_pass20}
\end{figure*}

\begin{figure*}
    \centering
    	\includegraphics[width=\linewidth]{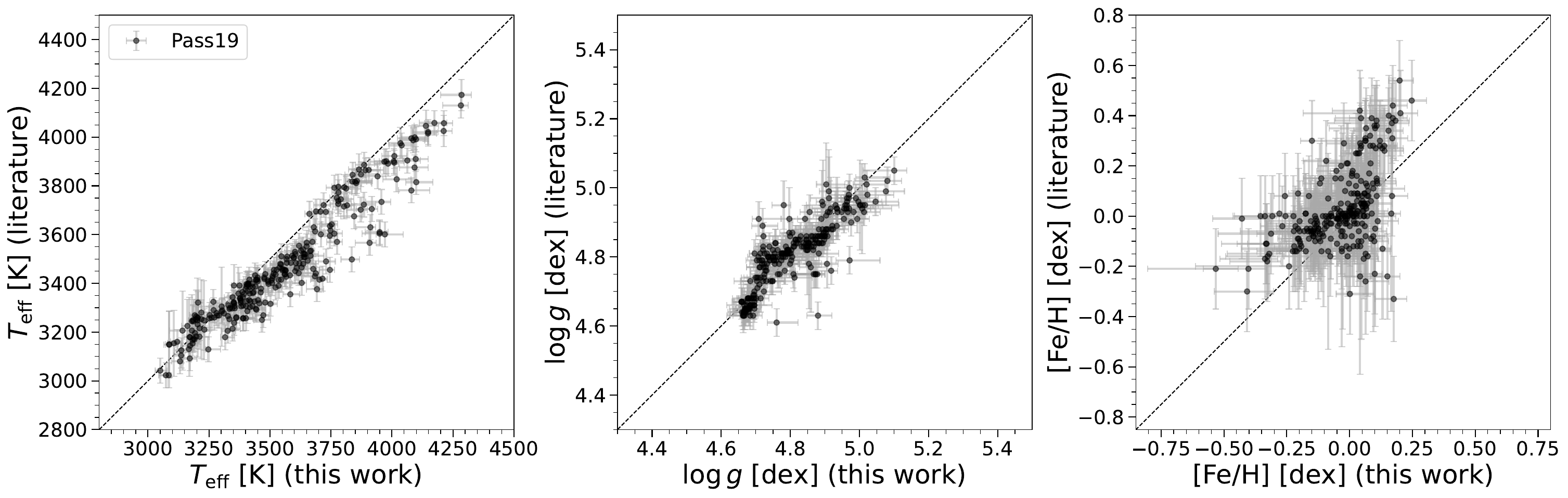}
    \caption{Comparison with \citetalias{pass2019}.}
    \label{fig:appendix_pass19}
\end{figure*}

\begin{figure*}
    \centering
    	\includegraphics[width=\linewidth]{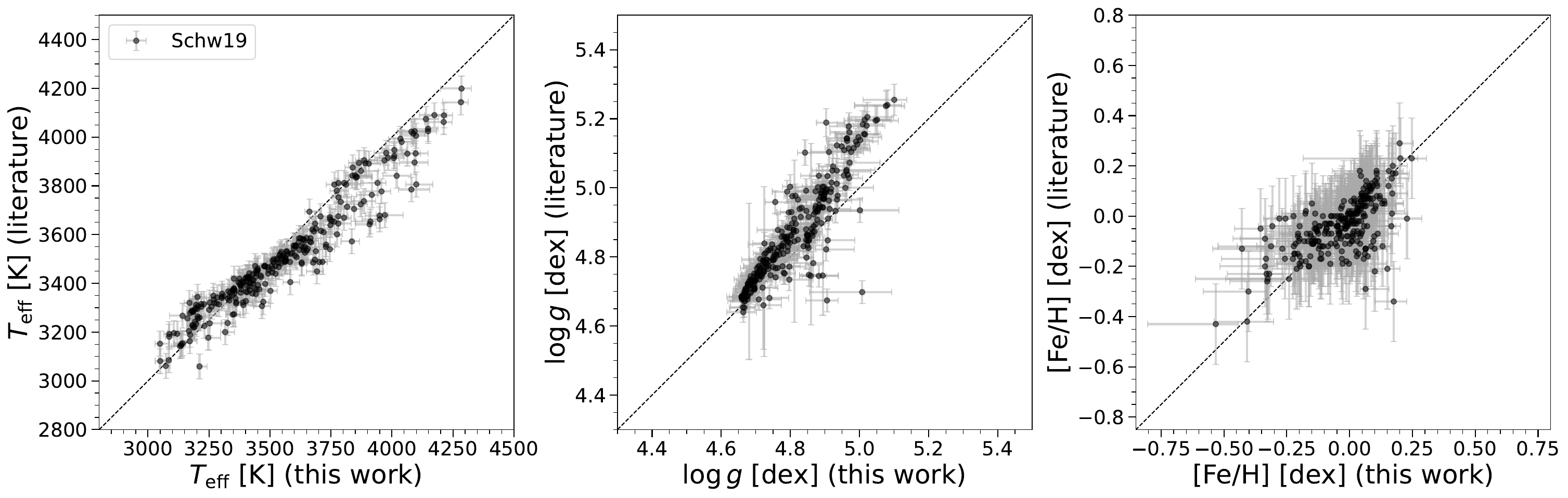}
    \caption{Comparison with \citetalias{schw19}.}
    \label{fig:appendix_schw19}
\end{figure*}

\begin{figure*}
    \centering
    	\includegraphics[width=\linewidth]{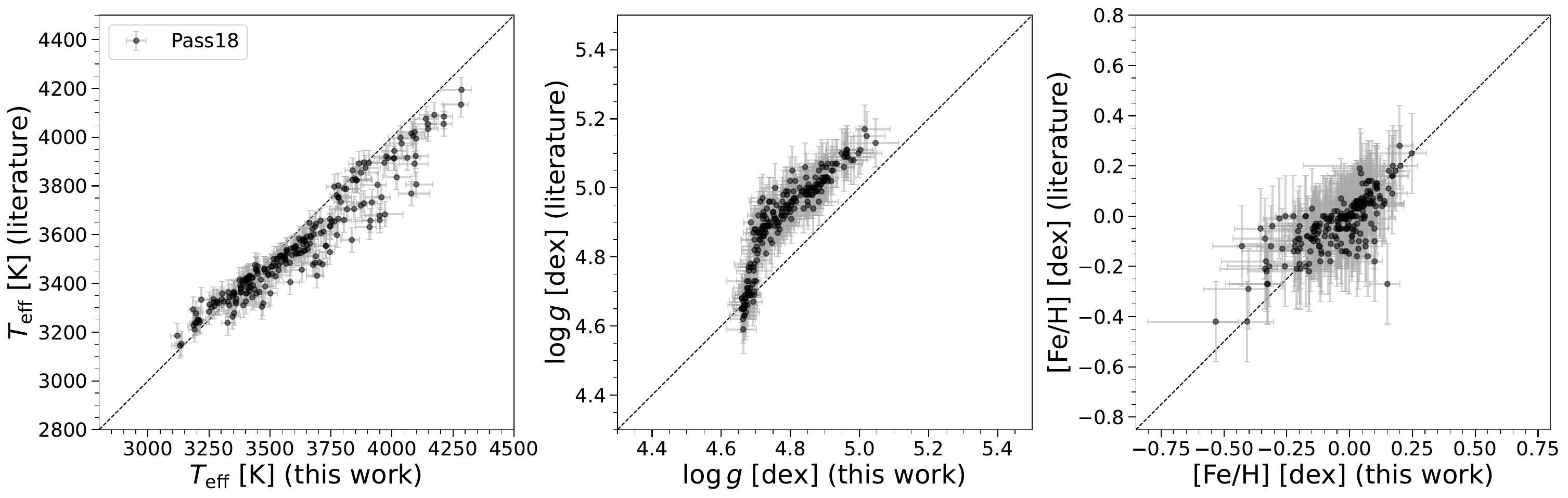}
    \caption{Comparison with \citet{pass18}.}
    \label{fig:appendix_pass18}
\end{figure*}

\begin{figure*}
    \centering
    	\includegraphics[width=0.68\linewidth]{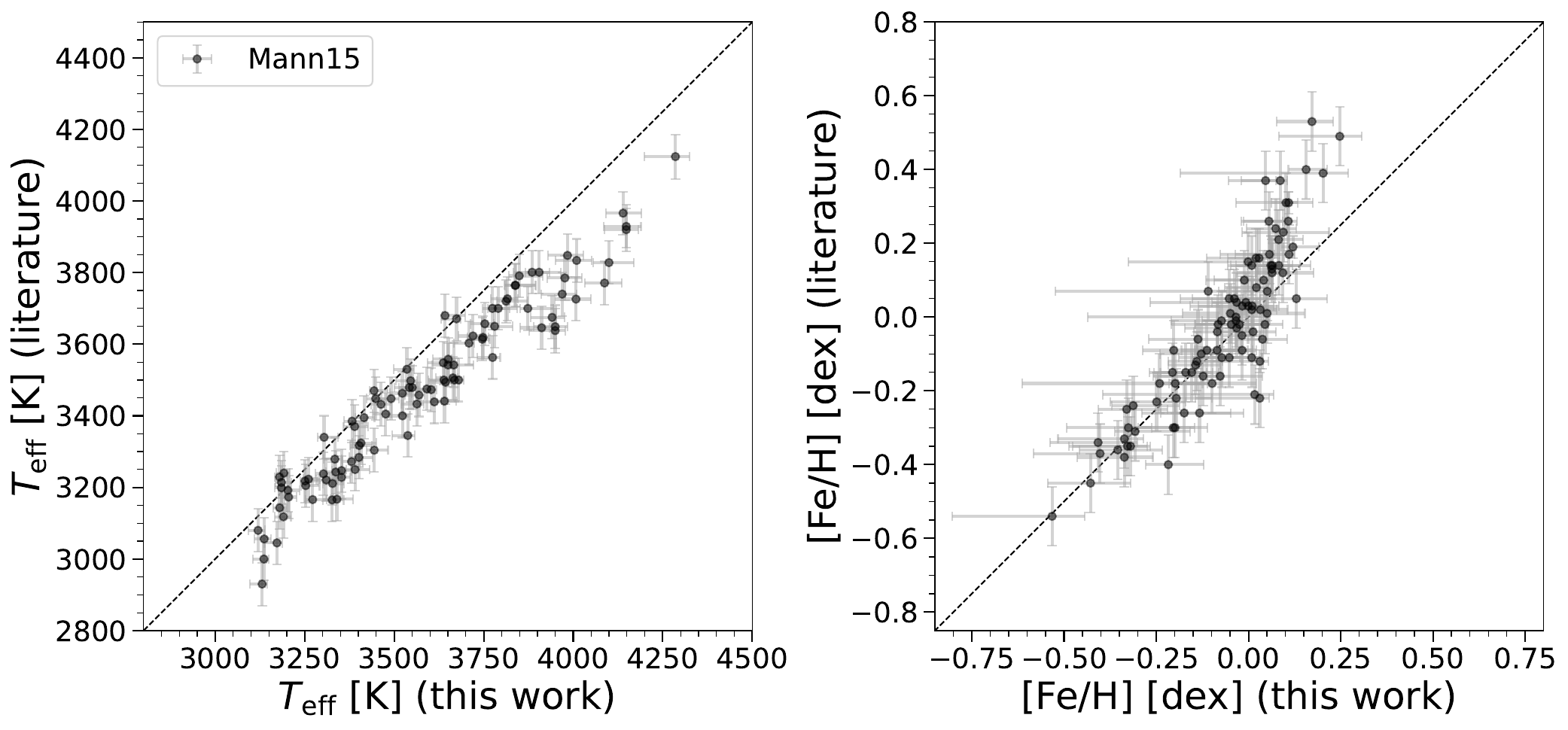}
    \caption{Comparison with \citet{mann2015}.}
    \label{fig:appendix_mann15}
\end{figure*}

\begin{figure*}
    \centering
    	\includegraphics[width=0.68\linewidth]{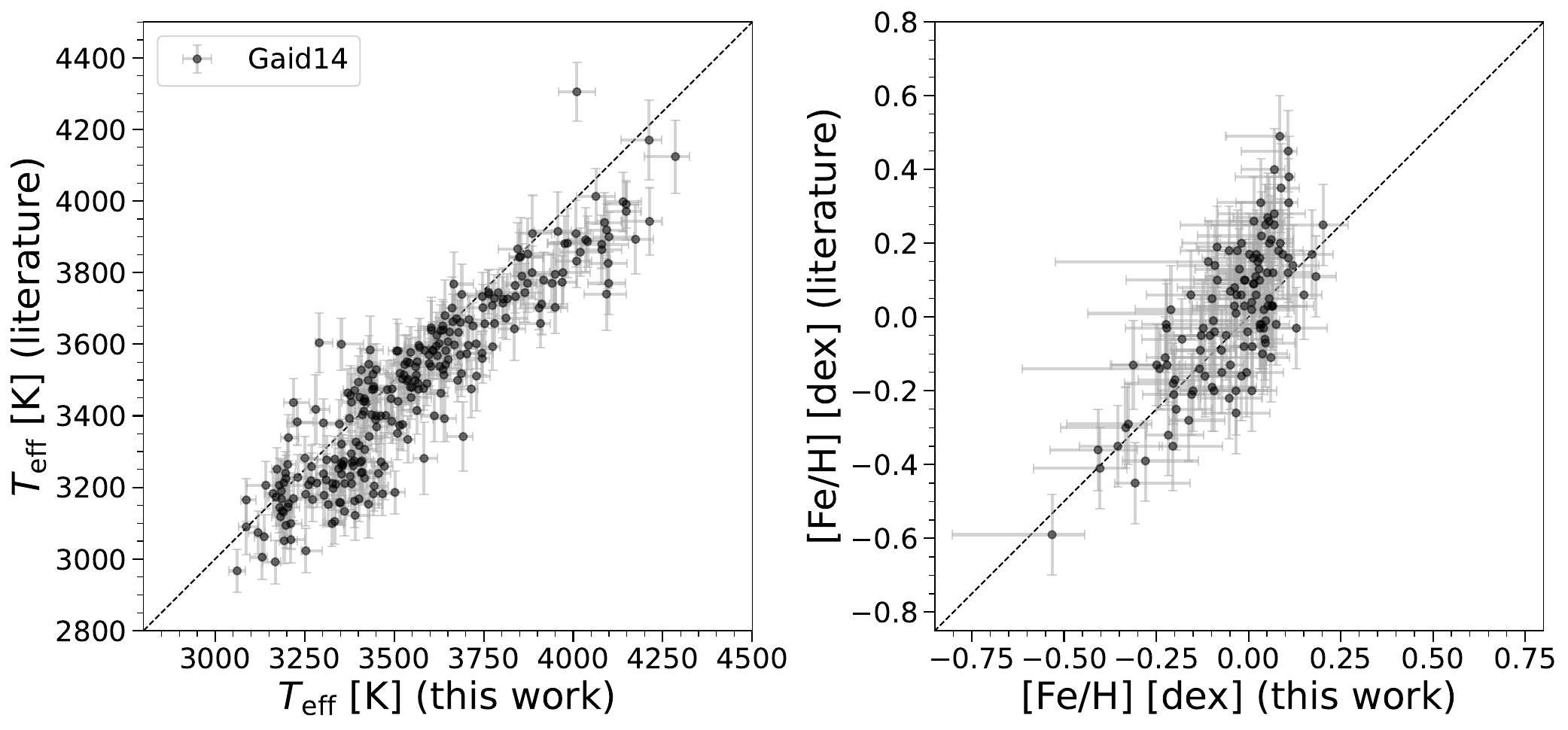}
    \caption{Comparison with \citet{gaid14}.}
    \label{fig:appendix_gaid14}
\end{figure*}

\begin{figure*}
    \centering
    	\includegraphics[width=0.68\linewidth]{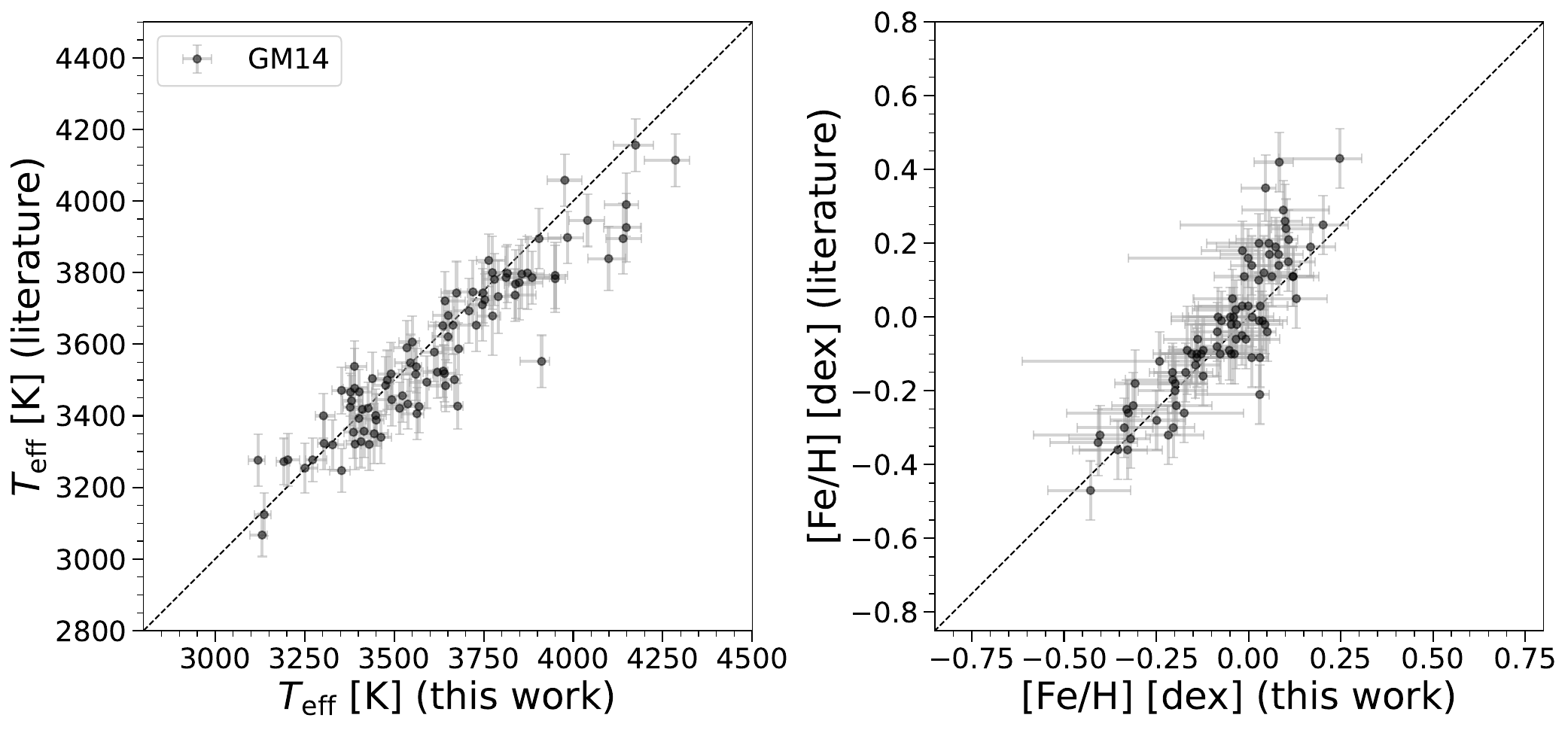}
    \caption{Comparison with \citet{GM14}.}
    \label{fig:appendix_gm14}
\end{figure*}

In this appendix, we provide an extensive comparison of this work with different results from the literature, as discussed in Section \ref{sec:lit_comp}. Also, we repeat the comparison shown in Figs. \ref{fig:scatter_teff}, \ref{fig:scatter_logg} and \ref{fig:scatter_mh}, but including the error bars. Table \ref{tab:comparison_lit_app} replicates Table \ref{tab:comparison_lit} for the additional literature collections. Figures \ref{fig:appendix_bello23}, \ref{fig:appendix_mar21}, \ref{fig:appendix_pass20}, \ref{fig:appendix_pass19}, \ref{fig:appendix_schw19}, \ref{fig:appendix_pass18}, \ref{fig:appendix_mann15}, \ref{fig:appendix_gaid14}, and \ref{fig:appendix_gm14} show the comparison with \citetalias{bello2023}, \citetalias{mar21}, \citetalias{pass20}, \citetalias{pass2019}, \citetalias{schw19}, \citet{pass18}, \citet{mann2015}, \citet{gaid14}, and \citet{GM14}, respectively.

\begin{table*}
 \caption{Comparison between our results and the additional literature collections. The structure is the same as in Table \ref{tab:comparison_lit}.}
 \label{tab:comparison_lit_app}
 \centering
 \begin{tabular}{l c c c c c c c c c}
 
  \hline\hline
  \noalign{\smallskip}

  Reference & \multicolumn{3}{c}{$T_{\rm eff}$\,[K]} & \multicolumn{3}{c}{log\,$g$\,[dex]} & \multicolumn{3}{c}{[Fe/H]\,[dex]}\\

    & $\overline{\Delta}$ & rmse & $r_{\rm p}$ & $\overline{\Delta}$ & rmse & $r_{\rm p}$ & $\overline{\Delta}$ & rmse & $r_{\rm p}$ \\
  
  \noalign{\smallskip}
  \hline
  \noalign{\smallskip}

  Pass18$^{\,(a)}$ & -59 & 98 & 0.96 & 0.12 & 0.14 & 0.89 & 0.01 & 0.09 & 0.73 \\    
  Mann15$^{\,(b)}$ & -109 & 136 & 0.96 & \ldots & \ldots & \ldots & 0.04 & 0.11 & 0.89 \\  
  Gaid14$^{\,(c)}$ & -69 & 151 & 0.87 & \ldots & \ldots & \ldots & 0.05 & 0.14 & 0.75 \\  
  GM14$^{\,(d)}$ & -42 & 102 & 0.93 & \ldots & \ldots & \ldots & 0.04 & 0.10 & 0.88 \\
  
  \noalign{\smallskip}
  \hline
 \end{tabular}
 \tablefoot{$^{(a)}$ From \citet{pass18}. $^{(b)}$ From \citet{mann2015}. $^{(c)}$ From \citet{gaid14}. $^{(d)}$ From \citet{GM14}.}
\end{table*}

\end{document}